\documentclass[12pt]{article}

\usepackage{ifthen}

\usepackage{bbm}

\usepackage{graphicx} 

\usepackage{array}

\usepackage{float}

\usepackage{dsfont}

\usepackage{amstext,amsmath,amssymb}

\usepackage{rotating}

\usepackage{a4}

\usepackage{a4wide}

\usepackage{color}

\usepackage[colorlinks=true,linkcolor=red,citecolor=green,urlcolor=cyan,
bookmarks=true,bookmarksopen=true,pdfpagemode=None,pdfstartview=FitH]{hyperref}

\def\be{\begin{equation}}

\def\ee{\end{equation}}

\def\gs{\mathrel{

   \rlap{\raise 0.511ex \hbox{$>$}}{\lower 0.511ex \hbox{$\sim$}}}}

\def\ls{\mathrel{

   \rlap{\raise 0.511ex \hbox{$<$}}{\lower 0.511ex \hbox{$\sim$}}}}

\newcommand{\ba}{\begin{array}{c}}

\newcommand{\baz}{\begin{array}{cc}}

\newcommand{\bad}{\begin{array}{ccc}}

\newcommand{\bav}{\begin{array}{cccc}}

\newcommand{\baf}{\begin{array}{ccccc}}

\newcommand{\bea}{\begin{equation} \begin{array}{c}}

\newcommand{\eea}{ \end{array} \end{equation}}

\newcommand{\ea}{\end{array}}

\begin{document}

\title{
\vglue -0.3cm
\vskip 0.5cm
\Large \bf
An accurate analytic description of neutrino oscillations in matter} 
\author{
{E. Kh. Akhmedov$^{a,b}$\thanks{email: \tt akhmedov@mpi-hd.mpg.de}~~\,and
\vspace*{0.15cm} ~Viviana Niro$^a$\thanks{email: 
\tt viviana.niro@mpi-hd.mpg.de}
} \\ 
{\normalsize\em $^a$Max--Planck--Institut f\"ur Kernphysik, 
Postfach 103980} \\ {\normalsize\em D--69029 Heidelberg, Germany 
\vspace*{0.15cm}}
\\ 
{\normalsize\em $^{b}$National Research Centre Kurchatov 
\vspace*{-0.1cm}Institute}\\{\normalsize\em Moscow, Russia} 
}

\date{ \today}

\maketitle

\thispagestyle{empty}

\vspace{-0.8cm}

\begin{abstract}
\noindent  
A simple closed-form analytic expression for the probability of two-flavour 
neutrino oscillations in a matter with an arbitrary density profile is 
derived. Our formula is based on a perturbative expansion and allows  
an easy calculation of higher order corrections. The expansion parameter  
is small when the density changes relatively slowly along the neutrino path 
and/or neutrino energy is not very close to the Mikheyev-Smirnov-Wolfenstein 
(MSW) resonance energy. Our approximation is not equivalent to the adiabatic 
approximation and actually goes beyond it. 
We demonstrate the validity of 
our results using a few model density profiles, including the PREM density 
profile of the Earth.  It is shown that by combining the results obtained 
from the expansions valid below and above the MSW resonance one can obtain a 
very good description of neutrino oscillations in matter in the entire 
energy range, including the resonance region. 
\end{abstract}

\newpage


\section{\label{sec:intro}Introduction}
In most neutrino oscillation experiments neutrinos 
propagate substantial distances in matter before reaching a detector, and 
therefore an accurate description of neutrino oscillations in matter 
\cite{Wolfenstein:1977ue,Mikheev:1986gs} is an important ingredient of the 
analyses of the data. For a matter of an arbitrary density profile the 
neutrino evolution equation admits no closed-form solution, and one usually 
has to resort to numerical methods. While numerical integration of the 
evolution equation usually poses no problem, it is still highly desirable to 
have approximate analytic solutions, which may provide a significant insight 
into the physics of neutrino oscillations in matter, clarify the dependence 
of the oscillation probabilities on the neutrino parameters and in many cases 
help save the CPU time. To this end, a number of analytic solutions 
of the neutrino evolution equation in matter, 
based on various approximations, 
has been developed (for recent studies, see e.g. \cite{deHolanda:2004fd,Ioannisian:2004jk,Akhmedov:2004rq,Ioannisian:2004vv,
Akhmedov:2005yj,Liao:2007re,Supanitsky:2007ks,Ioannisian:2008ve}).

In this paper we derive a simple analytic expression for the two-flavour 
oscillation probability valid for an arbitrary matter density profile. 
We employ a perturbative approach  
based on the expansion in a parameter which is small when the density 
changes relatively slowly along the neutrino path and/or neutrino energy is 
not very close to the Mikheyev-Smirnov-Wolfenstein (MSW) 
\cite{Wolfenstein:1977ue,Mikheev:1986gs} resonance energy. Our approximation 
is not equivalent to the adiabatic approximation and actually goes beyond it. 
We demonstrate the validity of our results using a few model density profiles, 
including the important PREM profile \cite{Dziewonski:1981xy}, which gives 
a realistic description of matter density distribution inside the Earth. 
We also show that, by combining the results obtained for the energies below 
and above the MSW resonance ones, one can obtain an excellent description of 
neutrino oscillations in matter in the entire energy range. 
The simple form of our result and the wide range of its applicability are 
the two main advantages of this approach. 

An approach similar to ours has been employed in \cite{Akhmedov:1988kd,
Akhmedov:2005yj}. Unlike in those publications, in the present work 
we do not confine ourselves to the leading approximation, but also calculate 
the first and second order corrections and show that this improves the 
accuracy of the approximation drastically. 

The paper is organized as follows. In Sec.~\ref{sec:formula} we present the 
formalism used to derive our analytic solution. In Sec.~\ref{sec:Parab} we 
apply this method to the case of a parabolic and a power law matter 
potentials. In Sec.~\ref{sec:PREM} we present the results obtained in the case 
of the realistic PREM Earth's density profile. 
We discuss our results and conclude in Sec.~\ref{sec:End}.


\section{\label{sec:formula}The formalism}
In a number of important cases the full three-flavour neutrino oscillations 
can to a very good accuracy be reduced to effective two-flavour ones. These 
include $\nu_e \leftrightarrow \nu_\mu(\nu_\tau)$ oscillations either in the 
limit of vanishingly small 1-3 mixing, when the oscillations are essentially 
driven by the ``solar'' parameters $\Delta m_{21}^2$ and $\theta_{12}$, or  
at sufficiently high energies ($E\gtrsim 1$~GeV for oscillations in the 
Earth), when the 1-2 mixing in matter is strongly suppressed; in that case 
the oscillation probabilities are essentially independent of the ``solar'' 
parameters and are governed by $\Delta m_{31}^2$ and $\theta_{13}$.  
For definiteness, in our numerical examples we will concentrate on the 
second case, though our general discussion will be valid in both situations. 

Two-flavour oscillations of neutrinos in matter are described by the 
Schr\"odinger-like evolution equation \cite{Wolfenstein:1977ue,Mikheev:1986gs} 
\be
i\left(
\begin{array}{c}
\dot{\xi}\\
\dot{\eta}
\end{array}
\right)
=
\left(
\begin{array}{cc}
-A & B\\
 B & A
\end{array}
\right)
\left(
\begin{array}{c}
\xi\\
\eta
\end{array}
\right),
\label{eq:amplitudeeq}
\ee
where the overdot denotes the differentiation with respect to the coordinate, 
and  $\xi$ and $\eta$ are respectively the probability amplitudes to find 
$\nu_{e}$ and $\nu_{a}$, the latter being a linear combination of $\nu_\mu$ 
and $\nu_\tau$. In the limit when the 1-3 mixing vanishes, $\theta_{13}\to 0$, 
one has $\nu_{a}=\cos \theta_{23} \nu_{\mu}-\sin \theta_{23} \nu_{\tau}$, 
whereas in the situations when the solar parameters play practically no role 
(e.g. for oscillations of high-energy neutrinos in the Earth), 
$\nu_{a}=\sin \theta_{23} \nu_{\mu}+\cos \theta_{23} \nu_{\tau}$. The quantities $A$ and $B$ 
in Eq.~(\ref{eq:amplitudeeq}) are 
\begin{eqnarray}
B &=& \delta\,\sin2\theta_{0}\,, \nonumber \\
A(x)&=& \delta\,\cos2\theta_{0}-V(x)/2\,.
\label{eq:AB}
\end{eqnarray}
Here the function $A(x)$ depends on the electron number density $N_e(x)$ 
through the Wolfenstein potential $V(x)$ defined as
$$
V(x)=\sqrt{2}\,G_{\textrm{F}}\,N_{e}(x)\cong7.54\times 10^{-14}\,Y_{e}(x)
\,\rho(x) (\textrm{g}/\textrm{cm}^{3}) ~\textrm{eV}, 
$$
where $G_F$ is the Fermi constant, $\rho(x)$ is the mass density of matter and 
$Y_e(x)$ is the number of electrons per nucleon. 
The parameter $\delta$ is defined as   
$\delta\equiv\Delta m^{2}/4 E$, and $\theta_{0}$ is the relevant mixing 
angle in vacuum. 
In the limit $\theta_{13}\to 0$ one has $\Delta m^{2}=\Delta m_{21}^2$, 
$\theta_{0}=\theta_{12}$, and the $\nu_e \leftrightarrow \nu_\mu(\nu_\tau)$ 
oscillation probabilities are given by
\be
P(\nu_e\rightarrow \nu_\mu;x)\,=\,P(\nu_\mu\rightarrow \nu_e;x)\,=\,
\cos^{2}\theta_{23}\,P_2(x)\,,
\ee
\be
P(\nu_e\rightarrow \nu_\tau;x)\,=\,P(\nu_\tau\rightarrow \nu_e;x)\,=\,
\sin^{2}\theta_{23}\,P_2(x)\,.
\ee
Here $P_2(x)$ is the effective two-flavour oscillation probability: 
\be
P_{2}(x)\,=\,P(\nu_{e}\rightarrow \nu_{a};x)\,\equiv\,|\eta(x)|^{2}\,
\label{P2}
\ee
(we assume the initial conditions $\xi(0) = 1$, $\eta(0) = 0$).
For oscillations of high-energy neutrinos in the Earth one has $\Delta m^{2}
=\Delta m_{31}^2$, $\theta_{0}=\theta_{13}$, and the $\nu_e \leftrightarrow \nu_\mu(\nu_\tau)$ 
oscillation probabilities are 
\be
P(\nu_e\rightarrow \nu_\mu;x)\,=\,P(\nu_\mu\rightarrow \nu_e;x)\,=\,
\sin^{2}\theta_{23}\,P_{2}(x)\,,
\ee
\be
P(\nu_e\rightarrow \nu_\tau;x)\,=\,P(\nu_\tau\rightarrow \nu_e;x)\,=\,
\cos^{2}\theta_{23}\,P_{2}(x)\,,
\ee
where, as before, $P_2(x)$ is given by Eq.~(\ref{P2}). 

Differentiating Eq.~(\ref{eq:amplitudeeq}), one can find decoupled second 
order differential equations for $\xi(x)$ and $\eta(x)$ 
\cite{Ermilova:1986ph,Notzold:1987cq}. 
The equation for the transition amplitude $\eta(x)$ reads   
\be
\ddot{\eta}+(\omega^{2}+i\dot{A})\eta=0~,
\label{eq:exact}
\ee 
where we have defined the function $\omega(x)$ as
\be
\omega^2(x)=A^{2}(x)+B^{2}\,.
\label{eq:omega}
\ee
Note that the instantaneous eigenvalues of the effective Hamiltonian in 
Eq.~(\ref{eq:amplitudeeq}) are $\pm \omega(x)$. The equation for $\xi(x)$ 
differs from Eq.~(\ref{eq:exact}) by the sign of the $\dot{A}$ term.    

It will be convenient for our purposes to rewrite Eq.~(\ref{eq:exact}) in 
the following form:
\be
\ddot{\eta}+(\omega^{2}-i\dot{\omega})\eta=(-i\dot{\Delta})\eta\,,
\label{eq:exact2}
\ee
where we have introduced the notation 
\be
\dot{\Delta}\equiv\dot{A}+\dot{\omega}\,. 
\label{eq:ddot1}
\ee
Eq.~(\ref{eq:exact2}) cannot in general be solved exactly, but, as we 
shall see, it admits a simple perturbative solution. To show that, let us 
first notice that, for energies (or densities) above the MSW resonance one, 
the quantity $\dot{\Delta}$ on the right hand side of Eq.~(\ref{eq:exact2}) 
is small. Indeed, from Eqs.~(\ref{eq:omega}) and (\ref{eq:AB}) it follows 
that for $V/2 - \cos 2\theta_0\,\delta \gg \sin 2\theta_0\,\delta$ (i.e. 
for $-A\gg B$) one has $\dot{\omega}\simeq- \dot{A}$, so that $\dot{\Delta}
\simeq 0$. The smallness of the parameter $\dot{\Delta}$ allows one to solve 
Eq.~(\ref{eq:exact2}) perturbatively, order by order. Expanding in powers of 
$\dot{\Delta}$, we find the equation for the $n$th order transition amplitude 
$\eta_{n}$ (with $n > 0$):
\be
\ddot{\eta}_{n}+(\omega^{2}-i\dot{\omega})\eta_{n}=(-i\dot{\Delta})
\eta_{n-1}\,.
\label{eq:etan}
\ee
The zero order transition amplitude $\eta_0$ satisfies the equation with 
the vanishing right hand side: 
\be
\ddot{\eta}_0+(\omega^{2}-i\dot{\omega}) \eta_0=0\,.
\label{eq:eta0}
\ee
Its solution for an arbitrary functional dependence of $\omega(x)$ on 
the coordinate can be readily found by considering the quantity $X_0\equiv 
\dot{\eta_0}-i\omega\eta_0$, which, as follows from (\ref{eq:eta0}),  
satisfies the first-order equation $\dot{X}_0+i\omega X_0=0$. 
Taking into account that the initial conditions $\xi(0)=1$, $\eta(0)=0$ also 
imply, through Eq.~(\ref{eq:amplitudeeq}), $\dot{\eta}(0)=-iB$, one finds 
\be
\eta_{0}(x)=-i\,B\, e^{i \phi\left(x\right)}\int^{x}_{0}dx_1\, 
e^{-2 i \phi\left(x_1\right)}\,,
\label{eq:etazero}
\ee
where
\be
\phi(x)\equiv\int_0^x \omega(x')\,dx'\,.
\label{eq:phi}
\ee
This yields the zero-order solution for the two-flavour transition 
probability $P_2(x)$ \cite{Akhmedov:1988kd,Akhmedov:2005yj}:
\be
\left[P_{2}(x)\right]_0 \equiv \left|\eta_{0}(x)\right|^{2} = B^{2}~
\left|\int^{x}_{0} dx_{1}~e^{-2 i \phi\left(x_{1}\right)}\right|^{2}\,.
\label{eq:P20}
\ee 

Assuming that the amplitude $\eta_{n-1}(x)$ on the right hand side of 
Eq.~(\ref{eq:etan}) is known, one can solve it for $\eta_{n}$. To this end, 
we introduce the quantity 
\be
X_n = \dot{\eta}_n-i\omega \eta_n\,,
\label{eq:x}
\ee
in terms of which Eq.~(\ref{eq:etan}) can be rewritten as 
\be
\dot{X}_{n}+i\omega X_{n}=(-i\dot{\Delta})\eta_{n-1}\,.
\label{eq:xn}
\ee
This can now be solved by the standard methods. First, we find the 
general solution of the homogeneous equation 
\be
\dot{X}_{n}+i\omega X_{n}=0\,,
\label{eq:omog}
\ee
which gives 
\be
X_{n}(x)=F\,e^{-i \phi(x)} 
\label{eq:Xn}
\ee
with $F$ an integration constant. Next, the solution of the inhomogenous 
equation (\ref{eq:xn}) is found by allowing $F$ to depend on the coordinate 
$x$ and substituting Eq.~(\ref{eq:Xn}) back into Eq.~(\ref{eq:xn}).
Taking into account the initial condition $F(0)=\dot{\eta}(0)-i\omega(0)
\eta(0)=-i B$, one finds 
\be
F(x)=\int^{x}_{0}dx_{1}\,e^{i \phi\left(x_{1}\right)}\,\left(-i \dot{\Delta}
\left(x_{1}\right)\right)\,\eta_{n-1}\left(x_{1}\right)\,-\,i B\,.
\label{eq:Fx}
\ee
The solution for $X_{n}$ is now given by Eq.~(\ref{eq:Xn}) with $F$ 
replaced by $F(x)$ from Eq.~(\ref{eq:Fx}). Once $X_n$ is known, it is 
straightforward to solve Eq.~(\ref{eq:x}) for $\eta_{n}$. This yields 
\be
\eta_{n}(x)=e^{i \phi\left(x\right)}\int^{x}_{0}dx_{1}\,e^{-2 i \phi
\left(x_{1}\right)}\int^{x_{1}}_{0}dx_{2}\,e^{i \phi\left(x_{2}\right)}
\,\left(-i \dot{\Delta}\left(x_{2}\right)\right)\,\eta_{n-1}\left(x_{2} 
\right)\,+\,\eta_{0}(x)\,,
\label{eq:finalsolution}
\ee
where we have used Eq.~(\ref{eq:etazero}). The corresponding $n$th order 
effective two-flavour oscillation probability is then found as 
$[P_2(x)]_n=|\eta_n(x)|^2$. 

Eq.~(\ref{eq:finalsolution}) represents the main result of our paper. 
It gives an analytic expression for the oscillation amplitude in the $n$th 
order in perturbation theory in terms of the lower-order solutions 
$\eta_{n-1}$ and $\eta_{0}$. For our numerical illustrations we will 
consider the solutions with $n=0$, 1 and 2. 

Eq.~(\ref{eq:finalsolution}) has been derived under the assumption that 
$\dot{\Delta}$ is a small parameter. As we pointed out before, this is true 
for energies above the MSW resonance one. This means that the perturbative 
approach considered above should, in general, fail for energies below the 
MSW resonance one. However, a simple modification of the above procedure 
leads to a description of neutrino oscillations valid below the MSW 
resonance. In order to show this, let us, instead of casting 
Eq.~(\ref{eq:exact}) in the form (\ref{eq:exact2}), rewrite it as 
\be
\ddot{\eta}+(\omega^{2}+i\dot{\omega})\eta=(-i\dot{\Delta})\eta\,,
\label{eq:belowRes}
\ee
where $\dot{\Delta}$ is now defined as 
\be
\dot{\Delta}=\dot{A}-\dot{\omega}~.
\label{eq:ddot2}
\ee
For small vacuum mixing angles, this is a small parameter below the MSW 
resonance, since in that case $A\gg B$ and so $\dot{\omega}\simeq \dot{A}$. 
Therefore, we can proceed with the perturbative approach, as before. 
Comparing Eqs.~(\ref{eq:belowRes}) and (\ref{eq:ddot2}) 
with Eqs.~(\ref{eq:exact2}) and (\ref{eq:ddot1}) respectively, we see that 
the two pairs of equations differ only by the sign of $\omega(x)$. Therefore 
the solution of Eq.~(\ref{eq:belowRes}) can be obtained from 
Eq.~(\ref{eq:finalsolution}) by simply replacing $\omega(x)$ by $-\omega(x)$.
This will also change the values of the oscillation probabilities obtained in 
all orders in perturbation theory except for the zero-order probability which, 
as can be seen from (\ref{eq:P20}), is invariant with respect to the flip 
of the sign of $\omega(x)$. 
As we shall see, by combining the results valid above and 
below the MSW resonance one can obtain a very good description of 
neutrino oscillations in matter in the entire energy range. 

Let us now discuss the expansion parameter of our perturbative approach. 
We have found that the corrections to the zero order amplitude $\eta_{0}$ 
are proportional to $\dot{\Delta}=\dot{A}\pm\dot{\omega}$, where the upper 
and lower signs refer to the energies above and below the MSW resonance, 
respectively. These quantities can be expressed through the mixing angle 
in matter $\theta_m$:%
\footnote{Note that $\sin 2\theta_m=B/\omega$, $\cos 2\theta_m=A/\omega$.}
\be
\dot{\Delta}=\dot{A} \pm \dot{\omega}=-\frac{\dot{V}}{2}\left[1 \pm 
\cos 2\theta_{m}\right]\,.
\ee 
Far above the MSW resonance one has $\cos 2\theta_m\simeq -1$, whereas 
far below the resonance $\cos 2\theta_m\simeq \cos 2\theta_0$, which is 
close to 1 in the case of small vacuum mixing. This demonstrates the 
smallness of $\dot{\Delta}$ in its corresponding domains of validity. 
At the MSW resonance one has $\cos 2\theta_m=0$, and $\dot{\Delta}$ is 
only small if $\dot{V}$ is. 

An examination of Eq.~(\ref{eq:finalsolution}) shows that the expansion 
parameter of our perturbative approach is actually $\sim |\dot{\Delta}|/
\omega^2$ (see Eq.~(\ref{eq:phi})). In various energy domains we have    
\be
\frac{|\dot{\Delta}|}{\omega^2}=
\left\{
\begin{array}{ccccccc}
\frac{|\dot{A} - \dot{\omega}|}{\omega^2} &\simeq& 
\frac{|\dot{V}|}{2}\frac{s^{2}_{2}\delta^{2}}
{2~(c_{2}\delta -V/2)^{4}}&  &\textrm{if}~~~(c_{2}\delta-V/2) \gg s_{2} 
\delta& \textrm{(below the resonance)}&\\\\
\frac{|\dot{A} \pm \dot{\omega}|}{\omega^2} &\simeq& \frac{|\dot{V}|}{2
s_2^2\delta^2}&  &\textrm{if}~~~|c_{2}
\delta-V/2| \ll s_{2} \delta& \textrm{(near the resonance)}&\\\\
\frac{|\dot{A} + \dot{\omega}|}{\omega^2} &\simeq& 
\frac{|\dot{V}|}{2}\frac{s^{2}_{2}\delta^{2}}
{2\,(V/2 - c_{2}\delta)^{4}}&  &\textrm{if}~~~(V/2-c_{2}\delta) \gg s_{2} 
\delta& \textrm{(above the resonance)}&
\end{array}
\right.
\label{eq:exppar}
\ee
where we have used the shorthand notation $c_2\equiv \cos 2\theta_0$, 
$s_2\equiv \sin 2\theta_0$. From Eq.~(\ref{eq:exppar}) it is easy to see that outside the MSW 
resonance region the expansion parameter approximately satisfies 
\be
\frac{|\dot{\Delta}|}{\omega^2}\,\simeq \,\sin^2 2\theta_m\, \frac{|\dot{V}|}
{4\omega^2}\,=\,\sin 2\theta_m\, \gamma_{\rm MSW}^{-1}\,,
\label{eq:exppar2}
\ee
where 
$\gamma_{\rm MSW}=4\omega^3/(|\dot{V}|B)=4\omega^2/(|\dot{V}|\sin 2\theta_m)$ 
is the MSW adiabaticity parameter. 
Thus, for small mixing in matter ($\sin 2\theta_m\ll 1$) our 
approximation is better than the adiabatic one. Close to the resonance 
the two approaches have comparable accuracy. 


\section{\label{sec:Parab}
Two examples: parabolic and power law profiles} 
As a first study, we apply our formalism to two simple density distributions: 
a parabolic and a power law profile.

For the parabolic profile, we consider the following density distribution:
\be
\rho(x)=\rho_{0}\left[-~k~\frac{(x-L/2)^{2}}{L^{2}/4}+1\right]
\label{eq:parab}
\ee
with 
\be
\rho_{0}=\rho_{max}=8~ \rm{g/cm}^{3}\,,\qquad
k=
1-\frac{\rho_{min}}{\rho_{max}}=0.5\,,
\ee
and we take the baseline to be $L=10000$ km. Note that the parabolic density 
profile represents a good approximation for the density distribution felt by 
neutrinos in the Earth when they cross only the Earth's mantle.

Next, we analyze the case of the following power-law density distribution:
\be
\rho(x)=\rho_{0}\left(\frac{x_{0}}{x_{0}+x}\right)^{3}
\label{eq:power}
\ee
with
\be
x_{0}=10^{3}~\textrm{km}~~~~\textrm{and}~~~~\rho_{0}=10^{3}~\rm{g/cm}^{3}\,,
\ee
and we consider neutrino propagation over the distance $L=100$ km. The profile 
$\rho~\propto~x^{-3}$ represents a realistic description of the density 
distribution inside supernovae; note, however, that neutrino flavour 
transitions in supernovae are more adequately described by different methods 
(see, e.g., \cite{Dighe:1999bi}), and so we consider the profile 
(\ref{eq:power}) just for illustration. 

\begin{figure}[ht]
\begin{tabular}{cc}
\includegraphics[width=8cm,height=7cm]{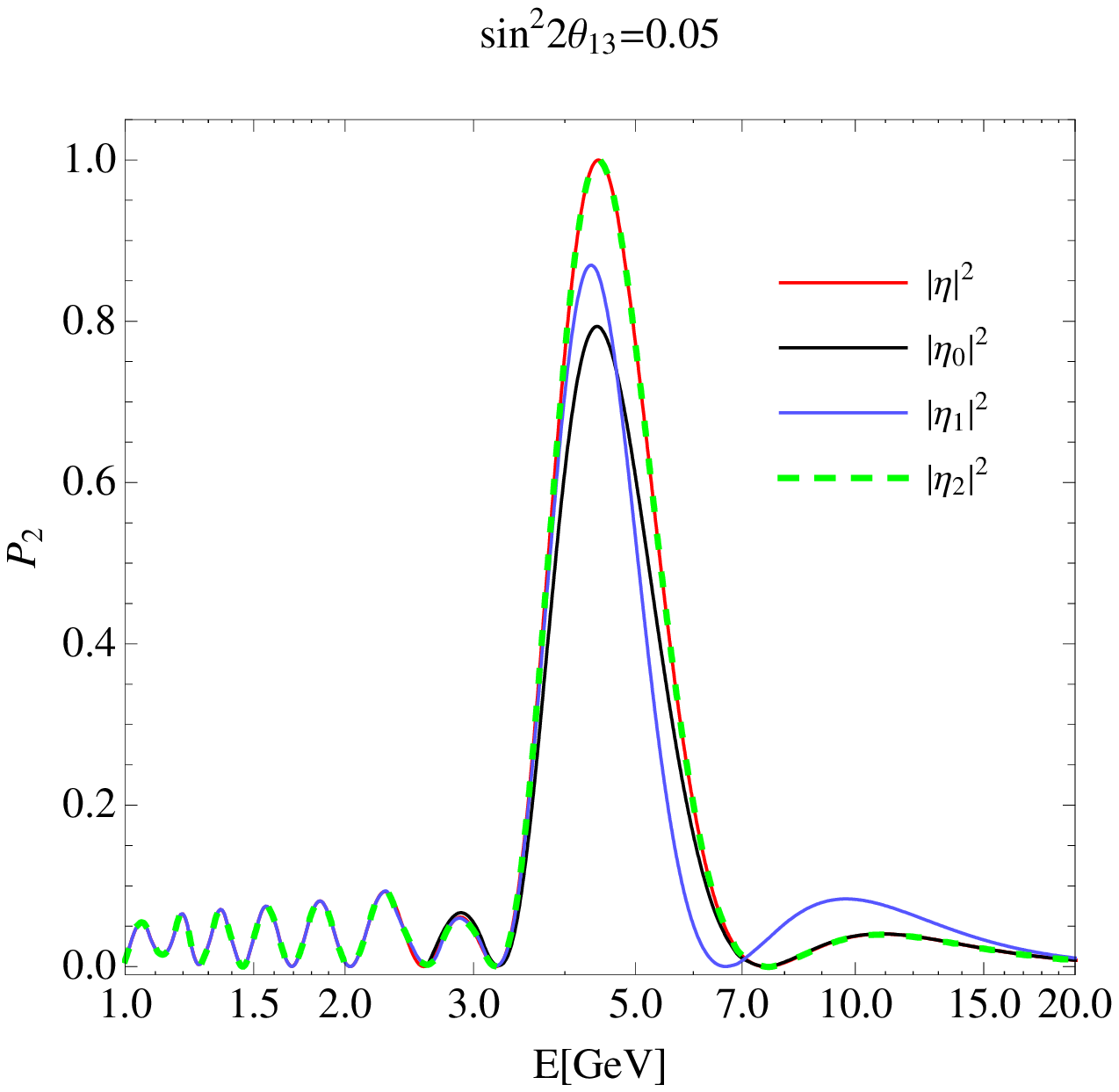}&
\includegraphics[width=8cm,height=7cm]{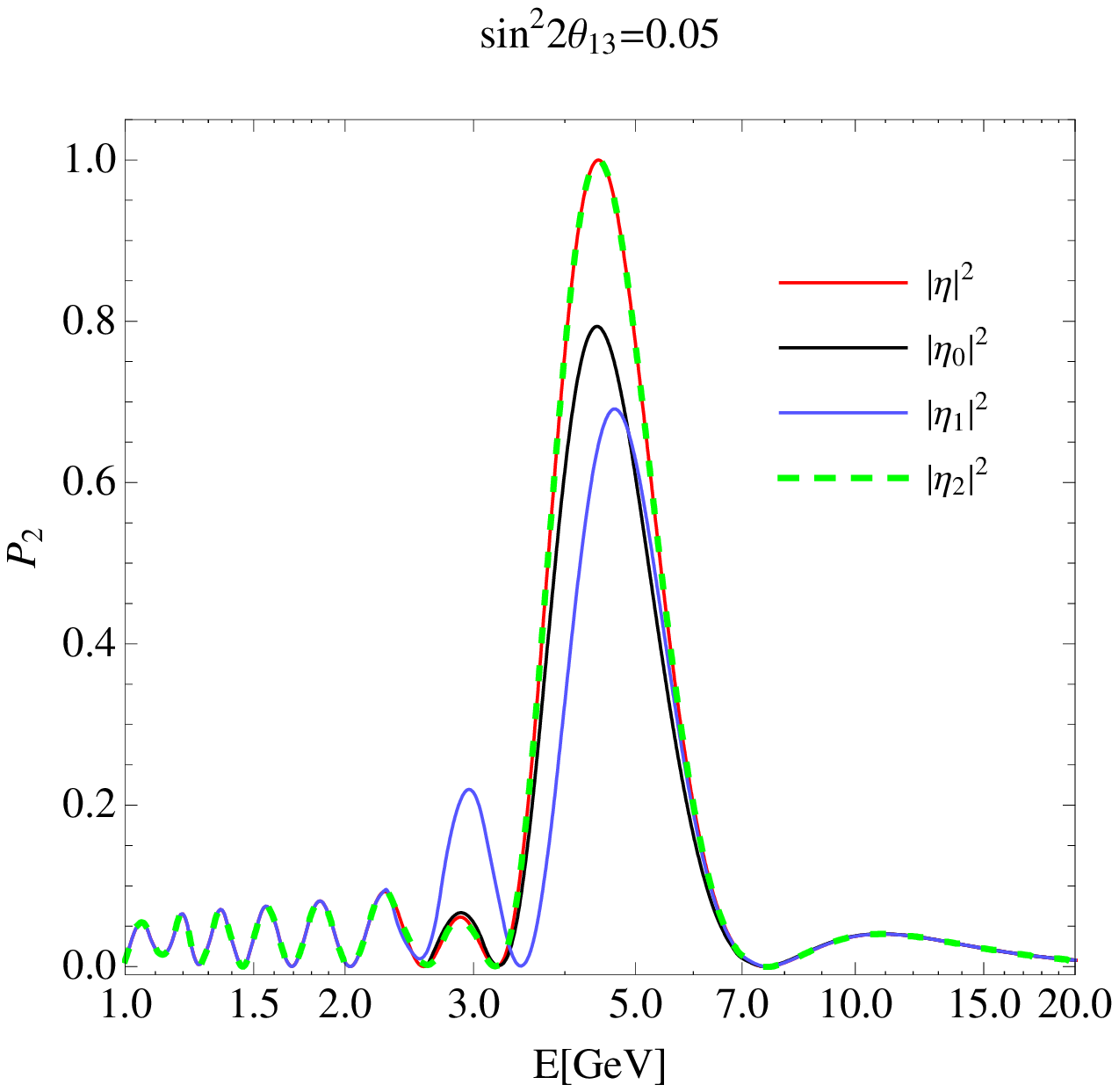}\\[1.5cm]
\includegraphics[width=7.5cm,height=7cm]{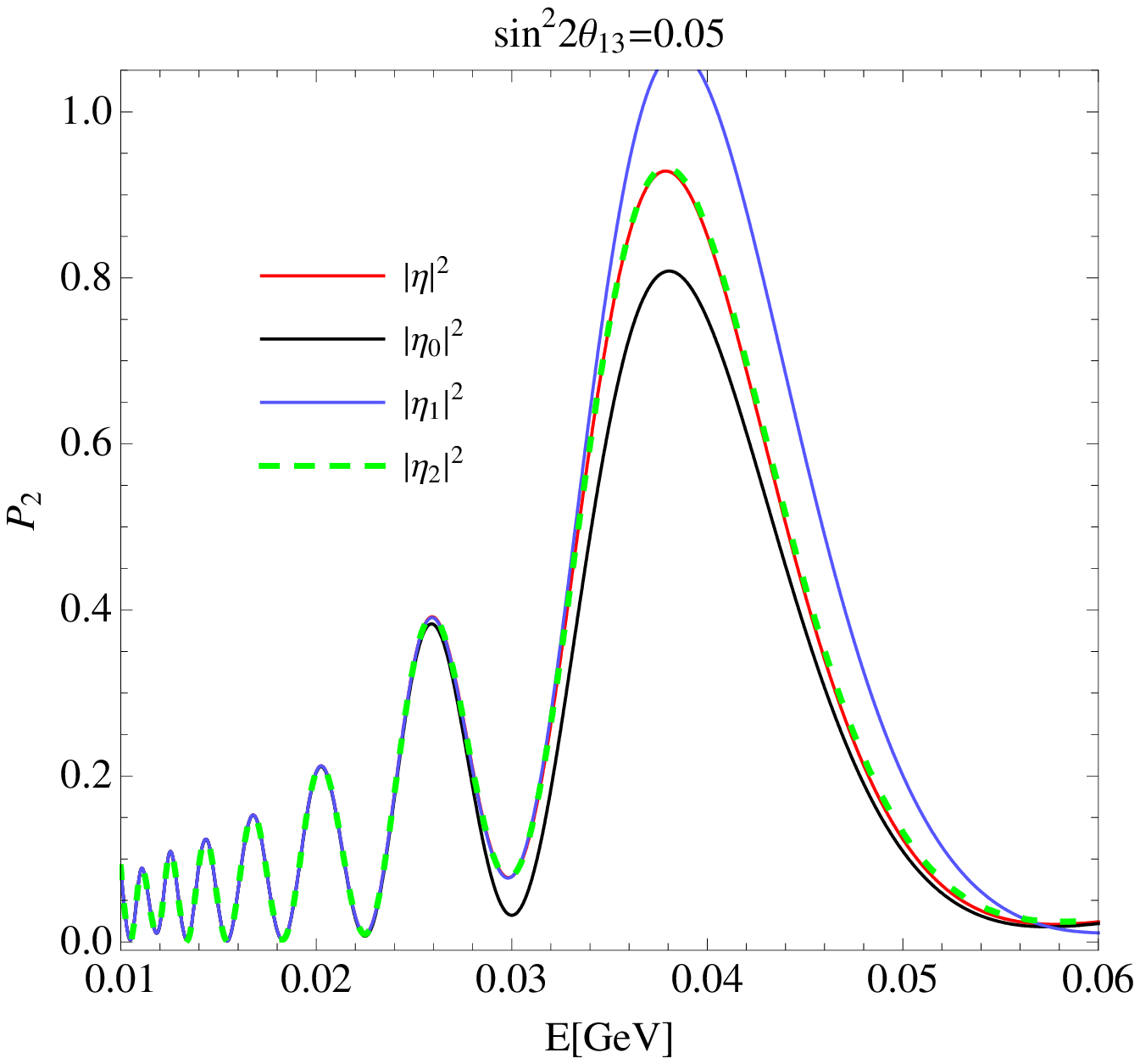}&
\includegraphics[width=7.5cm,height=7cm]{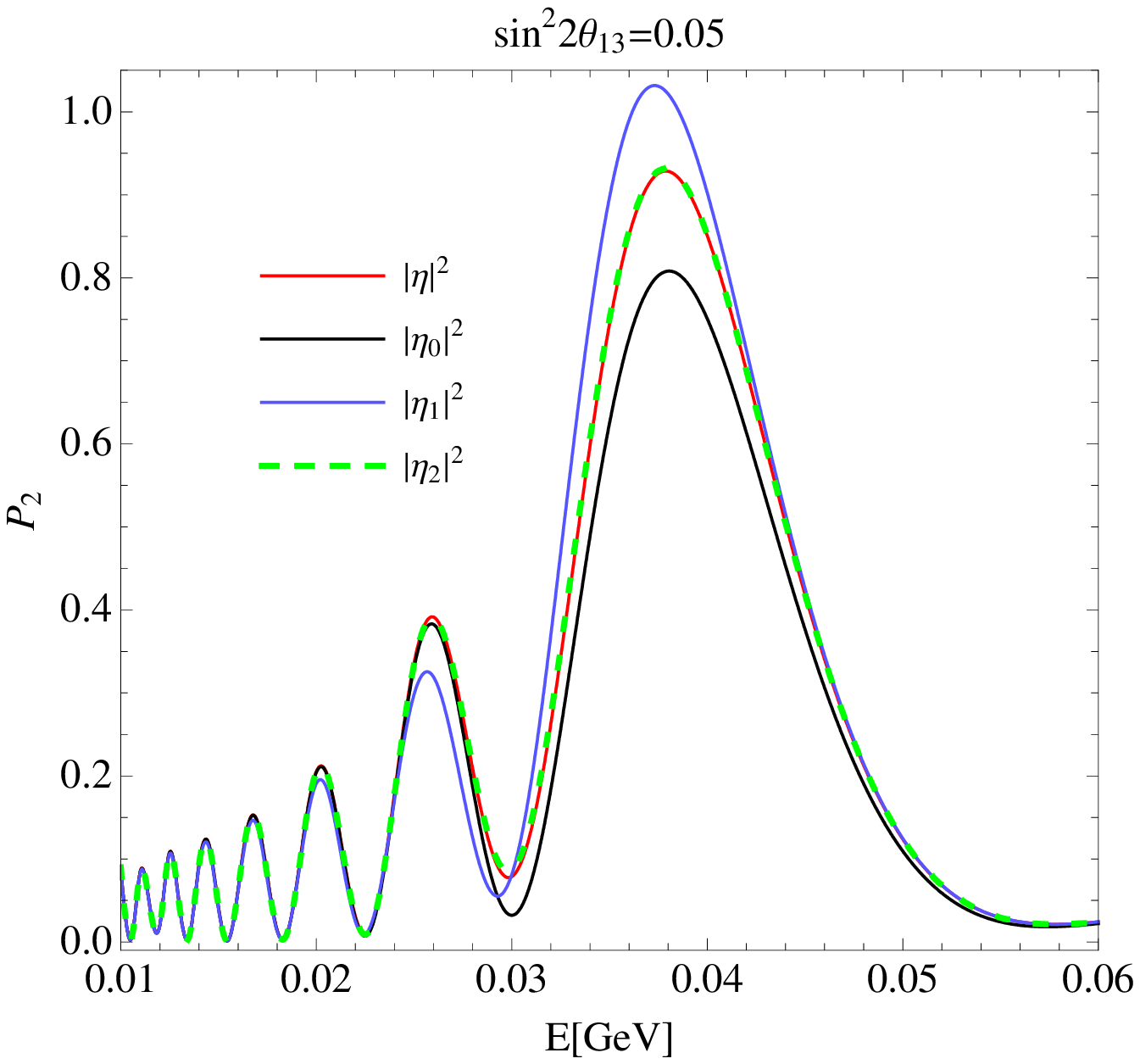}
\end{tabular}
\caption{\label{fig:parab}Oscillation probability $P_{2}$ versus 
neutrino energy $E$ in the case of the parabolic (upper plots) and 
power law (lower plots) density profiles. Left panels: probabilities 
obtained from the expansion valid below the MSW resonance, right 
panels: the same for the expansion valid above the resonance.  
We have taken $\Delta m^{2}=2.5~10^{-3}~$eV$^{2}$ and $Y_{e}=0.5$.}
\end{figure}

The results based on our perturbative analytic approach for the profiles 
(\ref{eq:parab}) and (\ref{eq:power}) are presented in Fig.~\ref{fig:parab}, 
where they are compared with the exact ones, obtained by direct numerical 
integration of the neutrino evolution equation~(\ref{eq:amplitudeeq}). 
The upper panels show the oscillation probabilities for the parabolic density 
profile and the lower ones, for the power-law profile (\ref{eq:power}). The 
left panels correspond to the expansion valid for energies below the MSW 
resonance ones, whereas the right panels were obtained for the expansion valid 
above the resonance energies.  
As expected, the zero-order approximation gives a good accuracy only 
outside the MSW resonance region (i.e., outside the intervals $E\sim 3$ 
-- 6 GeV for the parabolic profile and $E\sim$ 30 -- 50 MeV for the power-law 
one).%
\footnote{Note that, since the profiles (\ref{eq:parab}) and (\ref{eq:power})
(as well as the PREM profile considered in the next section) span a range of 
matter densities, neutrinos in an interval of energies experience the MSW 
resonance.} 
The first-order perturbative results obtained using the expansion valid 
below the MSW resonance extend slightly the region of good accuracy towards 
higher energies, closer to the MSW resonance, though in general fail for 
energies above the MSW resonance, whereas the first-order results found 
from the expansion valid above the MSW resonance extend the region of good 
accuracy to lower energies, but in general fail below the MSW resonance. Thus, 
the first-order calculation taken in their respective domains of applicability 
allow to achieve a good description of the exact results closer to the 
resonance energy than the zero-order solutions do, i.e. they reduce the 
energy domain in which the approximation fails. At the same time, as can 
be seen from Fig.~\ref{fig:parab}, the second-order probabilities $|\eta_2|^2$ 
practically coincide with the corresponding exact results, irrespectively of 
whether they are obtained using the expansion valid below or above the MSW 
resonance. 


\section{\label{sec:PREM}Propagation inside the Earth: PREM profile}
Neutrinos coming from various sources can propagate inside the Earth 
before reaching a detector. Examples are atmospheric neutrinos, neutrinos 
coming from WIMP annihilation inside the Earth or the Sun, as well as  
neutrinos studied in long-baseline accelerator experiments. We will consider 
here oscillations of high-energy neutrinos in the Earth, for 
which we take the matter density distribution as described by the PREM profile 
\cite{Dziewonski:1981xy} (Fig.~\ref{fig:PREMprofile}). Note that the PREM 
profile is symmetric with respect to the midpoint of the neutrino trajectory, 
and therefore the two-flavour transition amplitude $\eta(x)$ obtained as a 
solution of Eq.~(\ref{eq:amplitudeeq}) is pure imaginary due to the time 
reversal symmetry of the problem \cite{Akhmedov:2001kd}. 
 
\begin{figure}[ht]
\begin{tabular}{cc}
\includegraphics[width=8cm,height=7cm]{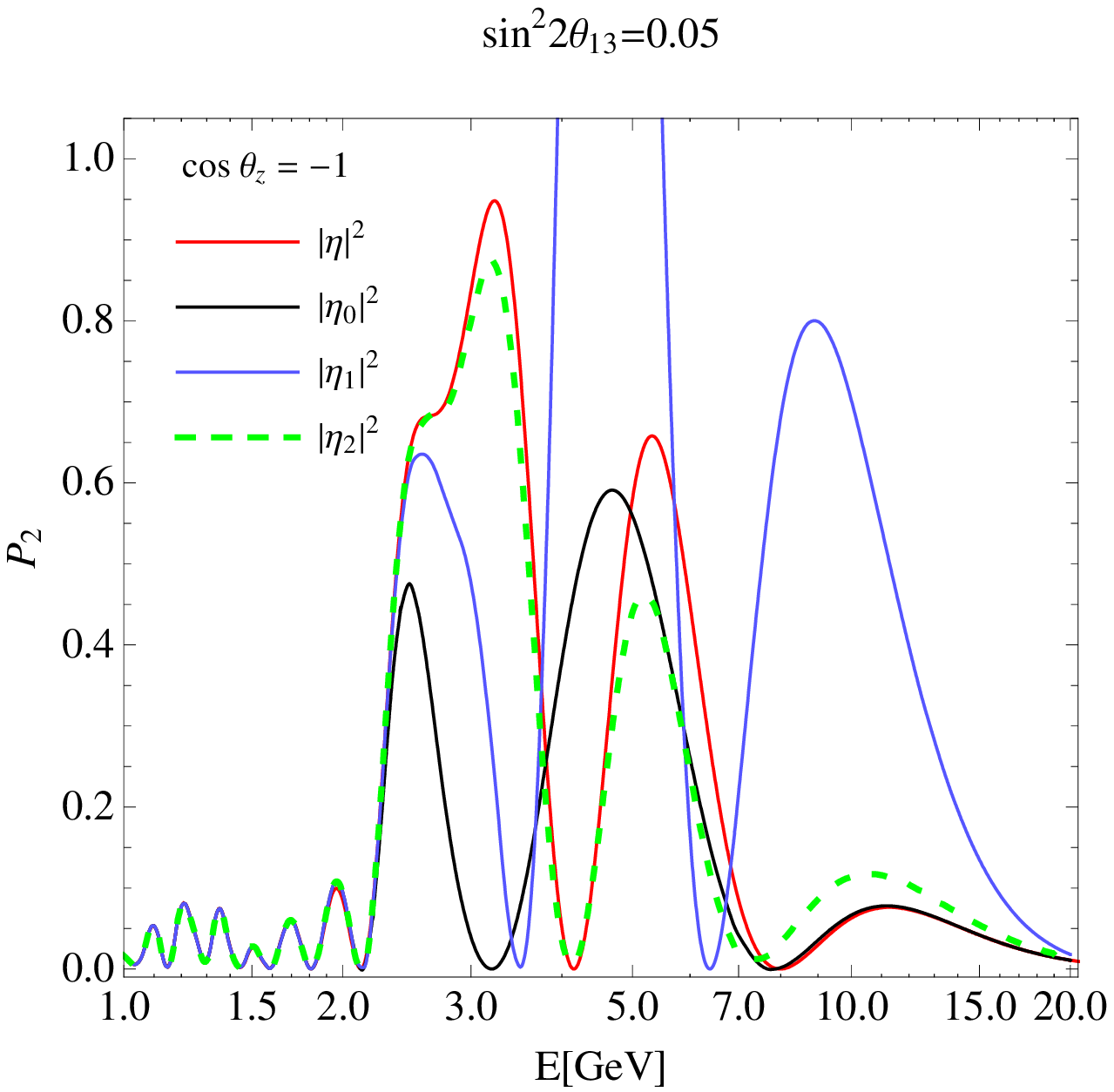}&
\includegraphics[width=8cm,height=7cm]{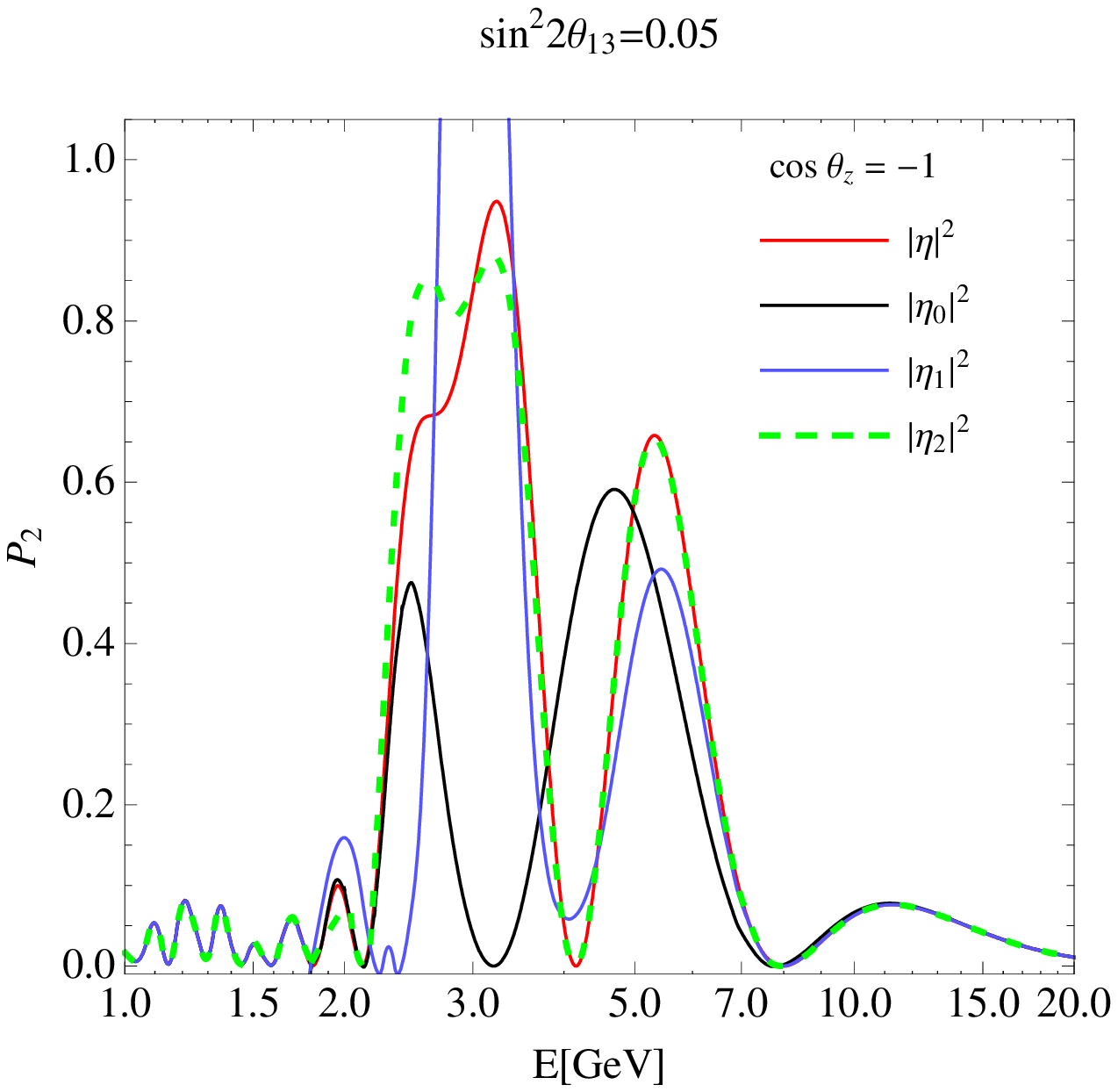}\\[1.5cm]
\includegraphics[width=8cm,height=7cm]{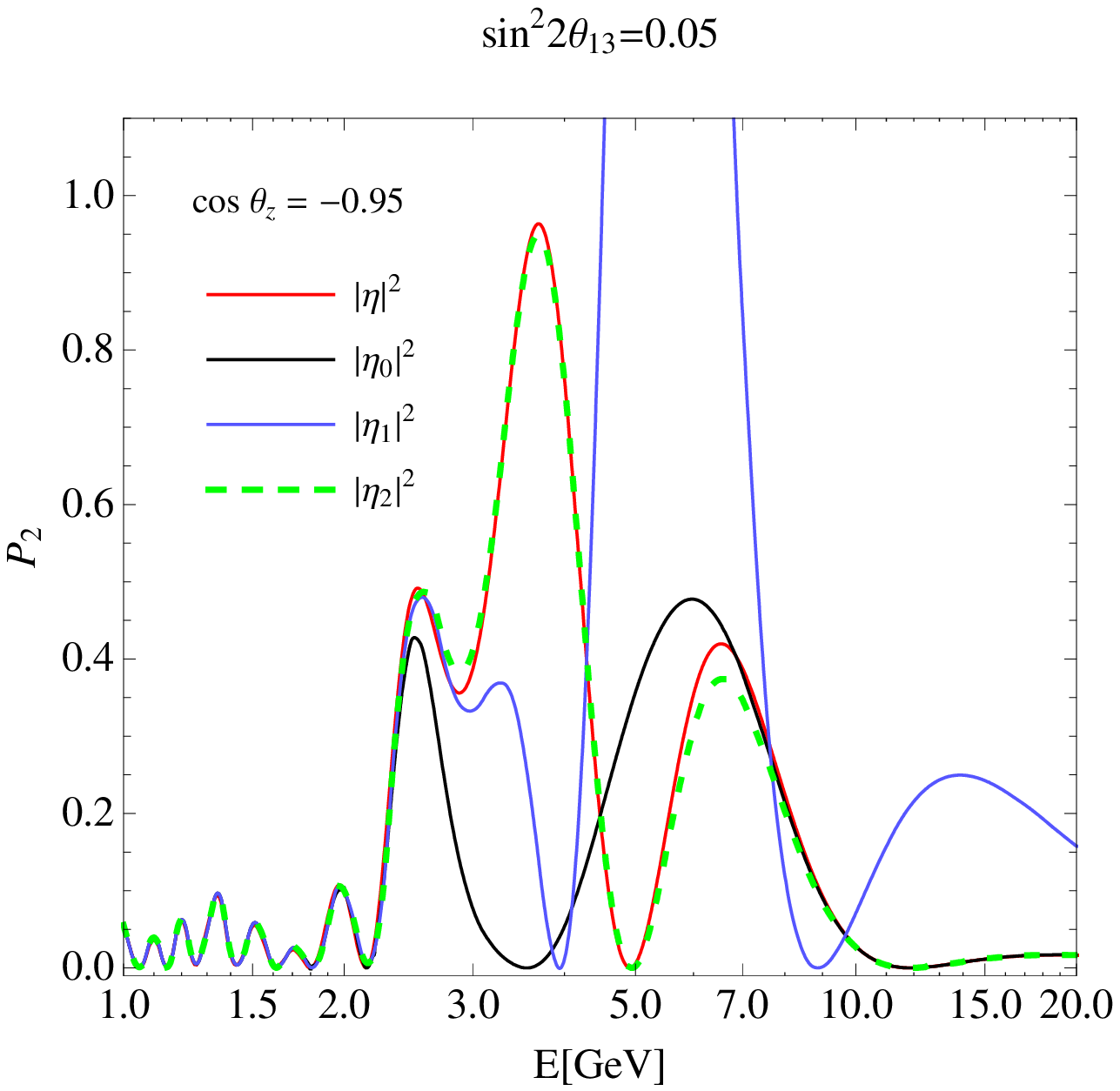}&
\includegraphics[width=8cm,height=7cm]{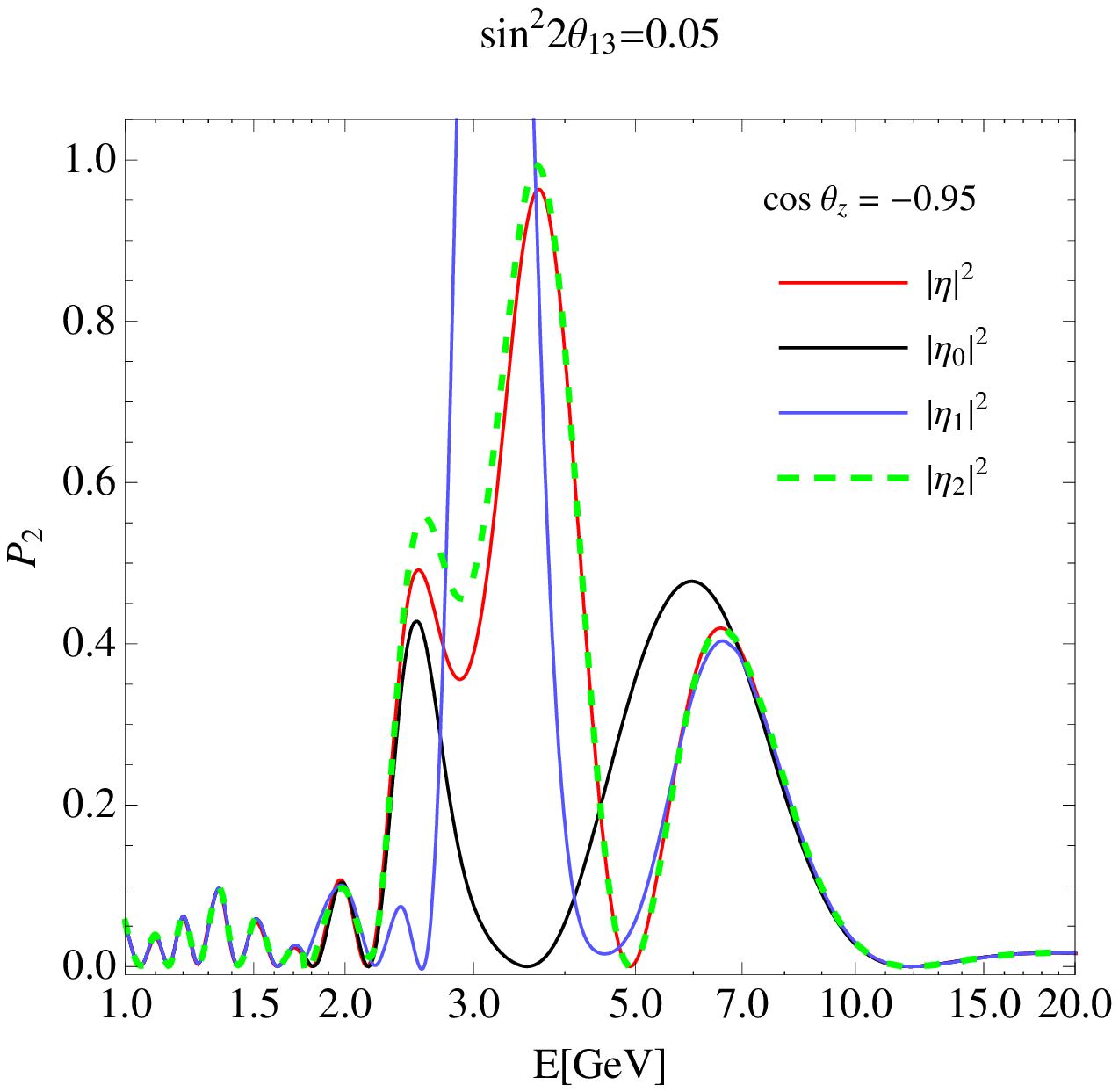}
\end{tabular}
\caption{\label{fig:FirstCorr}
Probability $P_{2}$ versus neutrino energy $E$ 
for neutrino oscillations in the Earth (PREM density profile) for two values 
of the zenith angle. Left panels: probabilities obtained from the expansion 
valid below the MSW resonance, right panels: the same for the expansion 
valid above the resonance. We have taken $\Delta 
m^{2}=2.5~10^{-3}~$eV$^{2}$.}
\end{figure}

In Fig.~\ref{fig:FirstCorr} we present the oscillation probability $P_2$ as 
a function of neutrino energy $E$ for two values of the zenith angle of the 
neutrino trajectory: $\cos \theta_{z}=-1$, when the neutrinos propagate the 
longest distance inside the Earth, traversing it along its diameter, and  
$\cos \theta_{z}=-0.95$, when they do not cross the inner core of the Earth. 
As in Fig.~\ref{fig:parab}, we compare the approximate solutions, up to the 
second order ($[P_2]_2=|\eta_{2}|^{2}$), with the exact solutions found 
by direct numerical integration of the neutrino evolution equation. In this 
figure (as well as in Figs.~\ref{fig:P2PHI} --~\ref{fig:sinvaluesbelow} below) 
in the left panels we present the oscillation probabilities obtained with 
the expansion valid below the MSW resonance energy, whereas the right 
panels show the results found from the expansion valid above the MSW 
resonance. 

It can be seen from Fig.~\ref{fig:FirstCorr} that the zero order probability 
$|\eta_{0}|^{2}$ reproduces accurately the exact one, $|\eta|^{2}$, only for 
energies that are outside the resonance region. Indeed, the 
two solutions nearly coincide for $E \le $ 2.5 GeV and for $E > 7$ GeV, 
but deviate substantially between these energies. As can be seen from the 
figure, the accuracy of the first order solutions is slightly better in 
their respective domain of validity: the solutions $|\eta_1|^2$ valid  
below the resonance (left panels of the figure) allow an accurate description 
of the probability for slightly higher energies than $|\eta_0|^2$ does, 
allowing to come closer to the MSW resonance from below; however, they  
fail badly (not even being bounded by 1) above the resonance. Likewise, the 
solutions $|\eta_1|^2$ valid above the resonance (right panels) allow 
to come closer to the MSW resonance from above, but fail below the resonance. 

At the same time, the second-order solution $|\eta_{2}|^{2}$ gives  
quite a good approximation to the exact probability $|\eta|^{2}$ for 
all energies, though the solutions obtained through the expansions in 
their corresponding domains of validity give a better accuracy in 
these energy domains. By combining the second-order solutions valid below 
and above the MSW resonance, one can obtain a very good description of the 
exact oscillation probability practically at all energies, including the 
resonance region. 
We have also checked that for trajectories that do not cross the core of 
the Earth ($\cos \theta_{z} >$  -0.838), for which the matter density 
profile seen by the neutrinos is relatively smooth, the second order 
solutions obtained through both expansions essentially coincide with the 
exact one for all energies. 

In Fig.~\ref{fig:P2PHI} we present the oscillation probability $P_{2}$, 
obtained in different orders in perturbation theory, as a function of the  
distance travelled by neutrinos inside the Earth for vertically up-going 
neutrinos ($\cos\theta_z=-1$) and two values of neutrino energy, 
$E=2.8$ GeV and 6 GeV. The figure clearly demonstrates how the accuracy 
improves with increasing order in perturbation theory; the second 
order solutions $|\eta_{2}|^{2}$ nearly concide with the exact probability 
$|\eta|^{2}$ along the entire neutrino path. 
 
Fig.~\ref{fig:Zenith3GeV} illustrates the dependence of the analytic 
solutions on the zenith angle for two values of the neutrino energy, $E=2.5$ 
GeV and 6 GeV. For both energies we show the solutions obtained using the 
expansions valid below and above the MSW resonance. The results agree with 
our expectations: the second order solution based on the expansion valid 
below the resonance reproduces the exact one extremely well for $E=2.5$ 
GeV but does not give a good accuracy (especially in the core region) 
for $E=6$ GeV, while the situation is opposite in the case of the solution 
corresponding to the expansion valid above the resonance.

Finally, in Fig.~\ref{fig:sinvaluesbelow} we show the dependence of the 
accuracy of the analytic solutions on the value of the vacuum mixing angle 
$\theta_0=\theta_{13}$. As one can see by comparing the upper panels with 
the corresponding lower ones, with decreasing value of $\theta_{13}$ the 
accuracy of our perturbative expansion improves. This is the consequence 
of the fact that the expansion parameter (\ref{eq:exppar2}) decreases with 
decreasing $\theta_{13}$.  


\section{\label{sec:End} Discussion and conclusions}
We have developed a perturbative approach for two-flavour neutrino 
oscillations in matter with an arbitrary density profile. The zero-order 
oscillation amplitude $\eta_0$ satisfies the equation which can be solved 
analytically for an arbitrary dependence of the matter density distribution on 
the coordinate along the neutrino path; higher order amplitudes are then 
obtained from the lower-order ones through a simple perturabative procedure. 
We have studied the zeroth, first and second order solutions and compared 
them with each other and with the exact oscillation probability 
obtained by numerical integration of the neutrino evolution equation. 
In all orders except the zeroth one, the expansion scheme 
depends on whether the neutrino energy is above or below the MSW 
resonance energy, and one has to consider these two cases separately. 

While the zero-order result gives a very good accuracy outside the resonance 
region, higher order corrections are necessary to achieve an accurate 
description of the oscillation probability in the vicinity of the MSW 
resonance. We have demonstrated how these corrections, when taken in 
their respective energy domains of validity, improve drastically    
the precision of the approximation. 

For the smooth density profiles that we have studied, we found that the second 
order oscillation probability reproduces the exact one extremely well in the 
whole interval of energies, including the MSW resonance region, independently 
of whether the expansion scheme valid below or above the resonance was used. 
The same is also true for the PREM density profiles in the case when neutrinos 
cross only the mantle of the Earth, since the density jumps experienced by 
neutrinos in that case are relatively small. The high accuracy of the second 
order approximation for smooth density profiles is related to the fact that 
our expansion parameter, Eq.~(\ref{eq:exppar2}), is proportional to 
$|\dot{V}|$. This parameter is smaller than the expansion parameter of the 
adiabatic approximation by the factor $\sin 2\theta_m$ and therefore our 
approach gives a better accuracy than the adiabatic expansion when the mixing 
in matter is small. Note that a different expansion of the same evolution 
equation (\ref{eq:exact}) was employed in \cite{Balantekin:1988aq}. 

\begin{figure}[ht]
\begin{tabular}{cc}
\includegraphics[width=7.2cm,height=6.2cm]{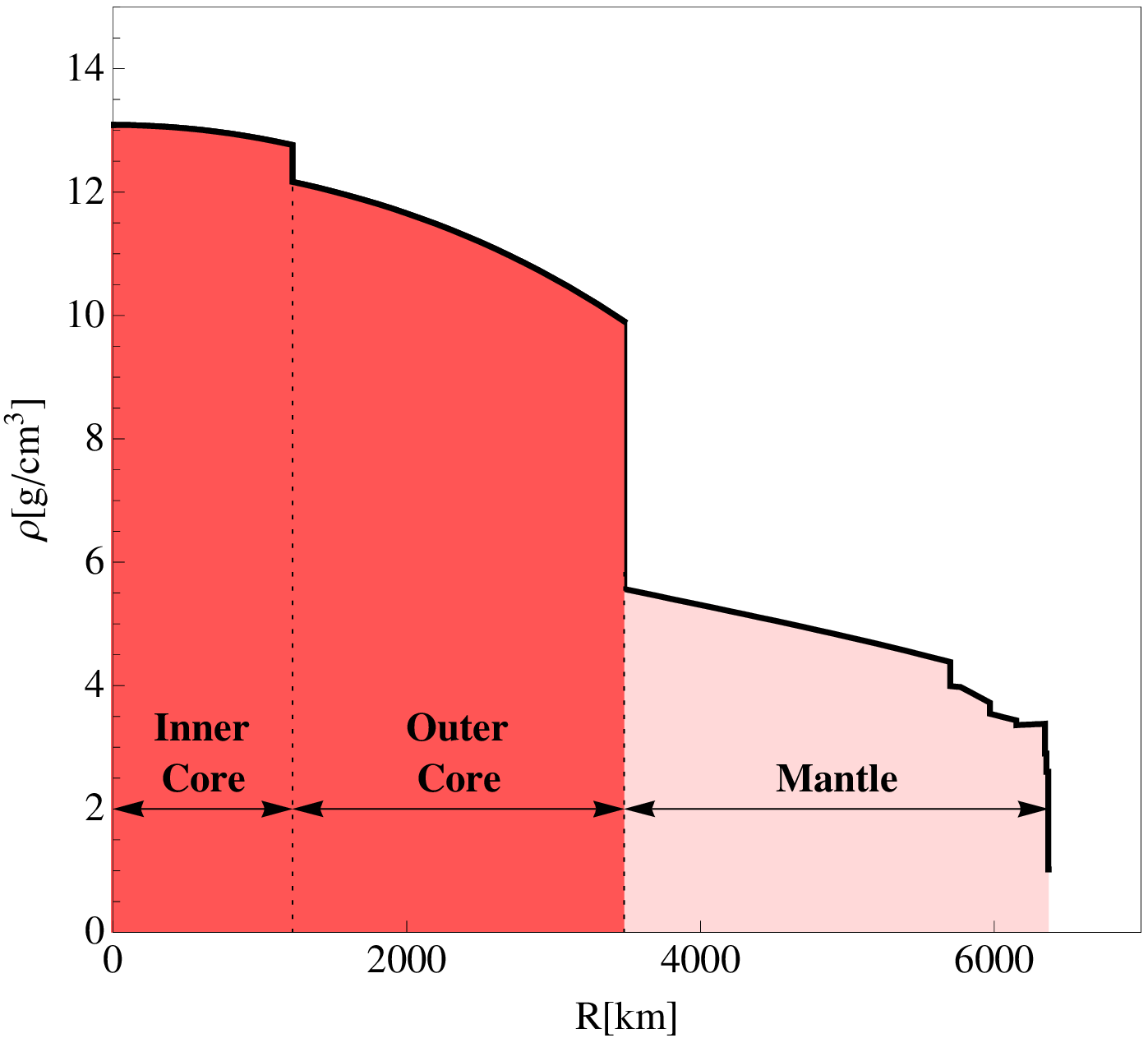}&
\includegraphics[width=7.2cm,height=6.2cm]{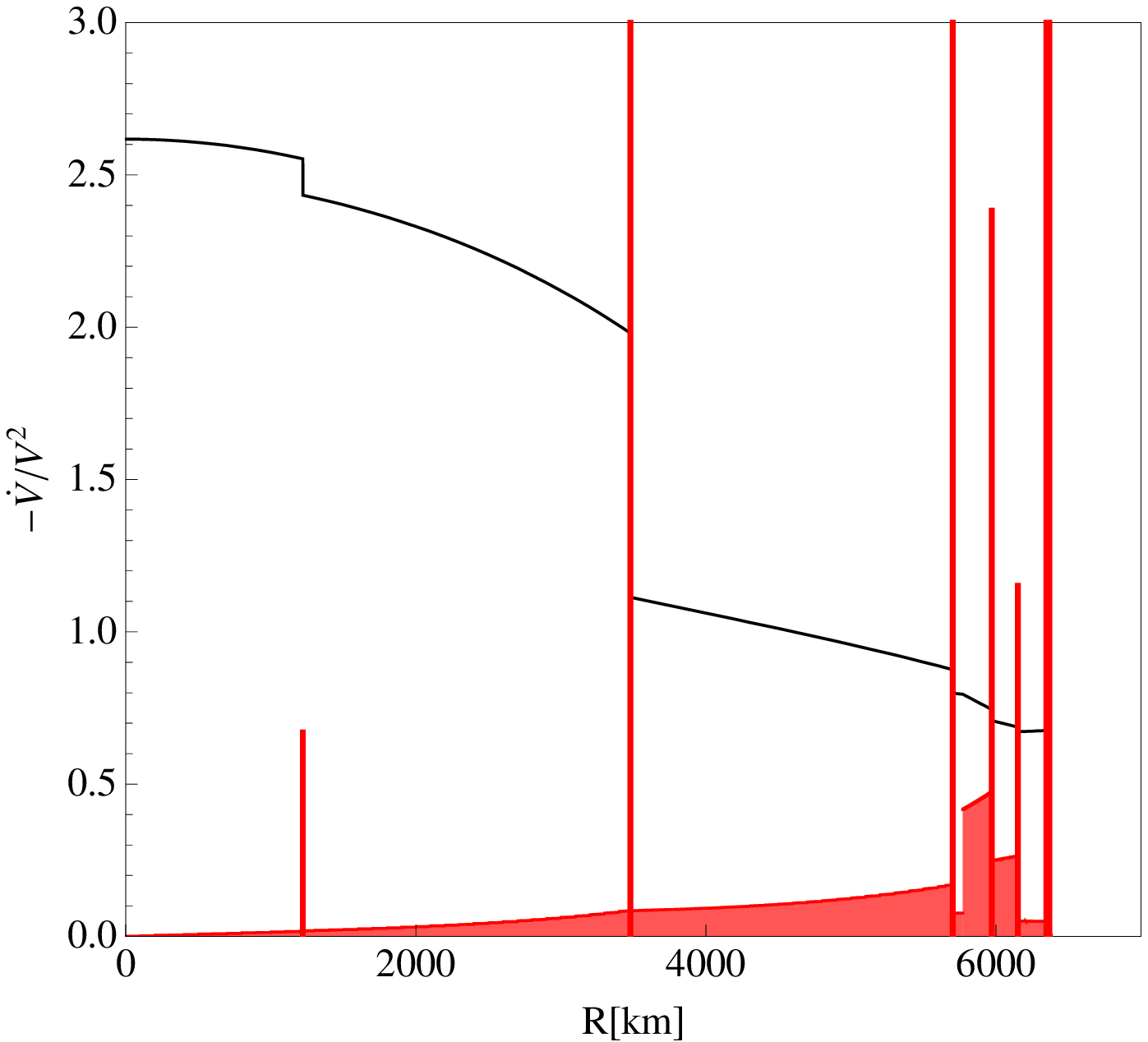}
\end{tabular}
\caption{\label{fig:PREMprofile}Left panel: matter density distribution 
inside the Earth as predicted by the PREM profile \cite{Dziewonski:1981xy}. 
Right panel: the function $-\dot{V}/V^{2}$ as calculated with the PREM 
profile with density jumps smoothed over the distance of 30 km. }
\end{figure}

For energies above the MSW resonance, our expansion parameter is 
essentially
\be
\frac{|\dot{\Delta}|}{\omega^2}\,\simeq \,
\sin^2 2\theta_m\,\frac{|\dot{V}|}{V^2}\,.
\label{eq:exppar3}
\ee
For the PREM density profile of the Earth, the function $|\dot{V}|/V^2$ is 
plotted in the right panel of Fig.~\ref{fig:PREMprofile}. As can be seen from 
the figure, in most of the coordinate space the value of this function does 
not exceed 
0.25. The spikes corresponding to the density jumps, though quite high, are 
very narrow; they do not destroy our approximation because their contributions 
get suppressed due to the integrations involved in the calculation of the 
higher-order corrections to the oscillation amplitude 
(see Eq.~(\ref{eq:finalsolution})). Still, these contributions are not 
negligible, especially for neutrinos crossing the Earth's core. As a result, 
for core-crossing neutrinos with energies close to the MSW resonance ones, 
even the second-order  oscillation probabilities are only adequate when taken 
in their respective energy domains of validity. By combining the solutions 
valid below and above the MSW resonance one obtains a very accurate 
description of neutrino oscillations in matter in the entire energy range. 

To conclude, we have derived a simple closed-form analytic expression for the 
probability of two-flavour neutrino oscillations in a matter with an arbitrary 
density profile. Our formula is based on a perturbative expansion and allows 
an easy calculation of higher order corrections. We have applied our formalism 
to a number of density distributions, including the PREM density profile of 
the Earth, and demonstrated that the second-order approximation gives a 
very good accuracy in the entire energy interval.    


\vspace{0.3cm}
\begin{center}
{\bf Acknowledgments}
\end{center}
We thank Andreas Hohenegger for the help with numerical calculations and 
Michele Maltoni for useful discussions.


\newcommand{\eprint}[1]{
arXiv: \href{http://arxiv.org/abs/#1}{\texttt{#1}}
}

\bibliographystyle{./apsrev}

\bibliography{nupr2}


\newpage

\newpage

\newpage

\begin{figure}[ht]
\begin{tabular}{cc}
\includegraphics[width=7.5cm,height=6.5cm]{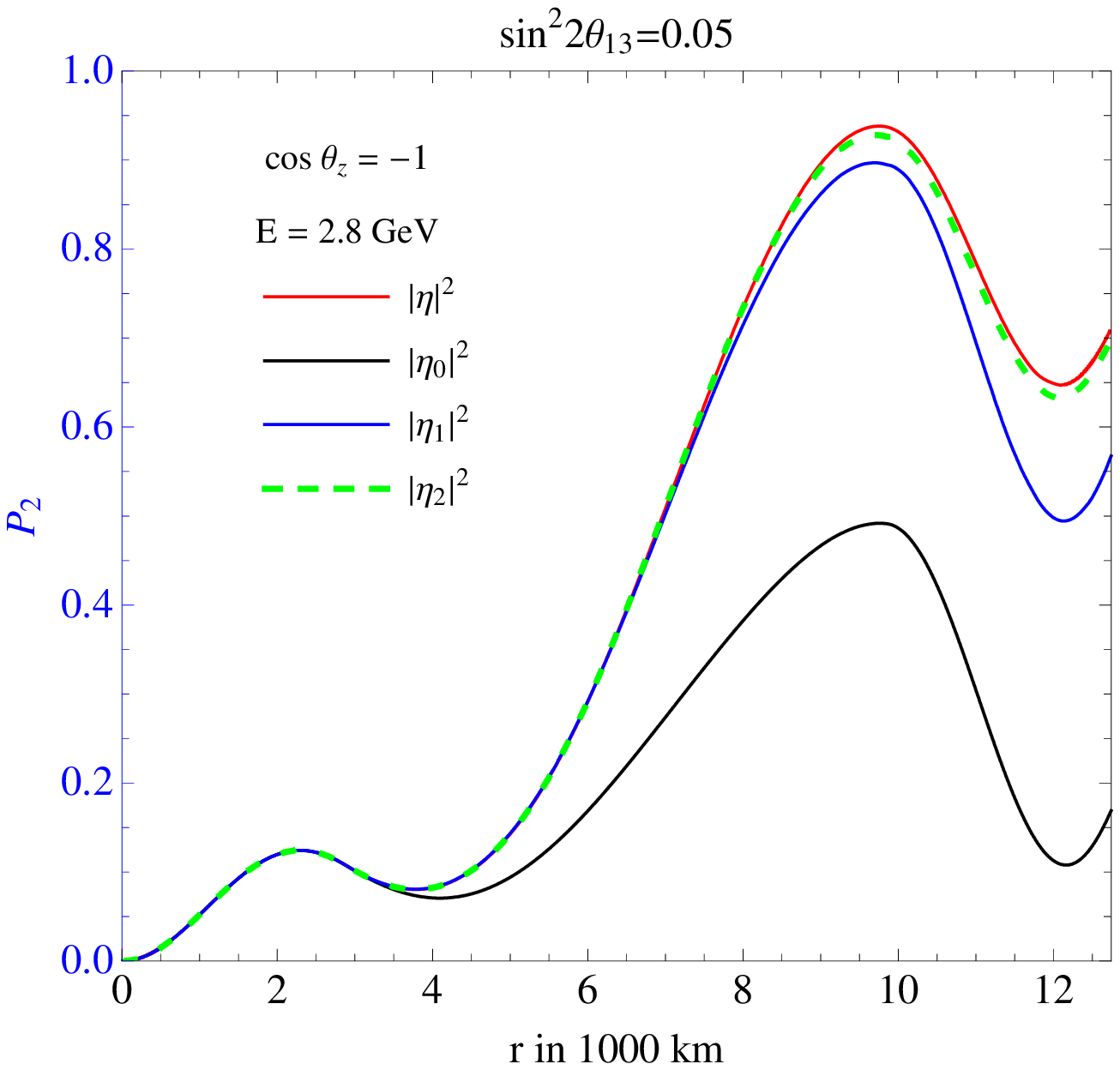}&
\includegraphics[width=7.5cm,height=6.5cm]{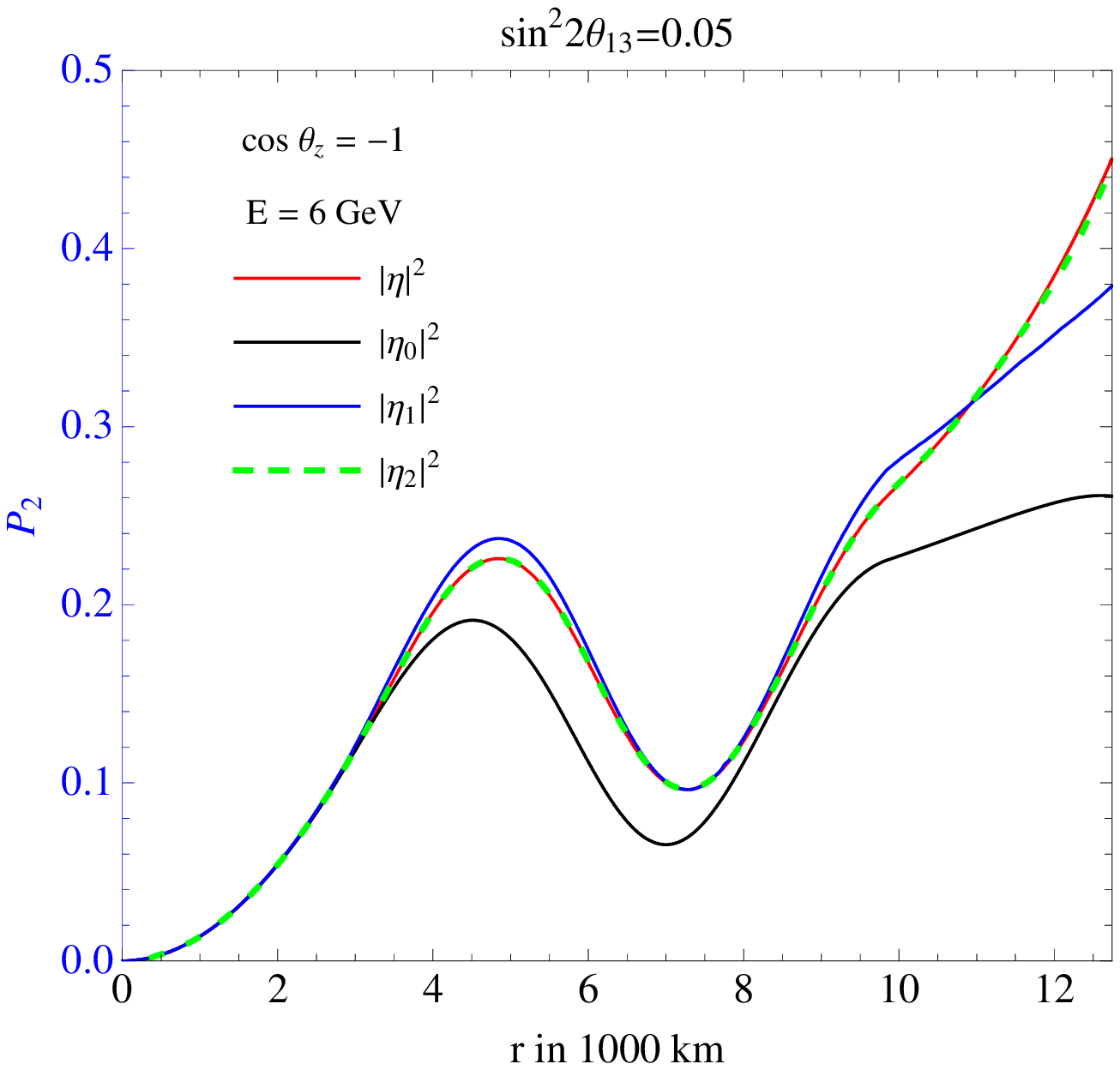}
\end{tabular}
\caption{\label{fig:P2PHI}Oscillation probability $P_{2}$ in different 
orders in perturbation theory versus the distance travelled by neutrinos 
inside the Earth, for $\cos \theta_{z}$ = -1.0 and for two values of 
neutrino energy ($E$= 2.8 GeV and 6 GeV). Left panels: probabilities 
obtained from the expansion valid below the MSW resonance, right 
panels: the same for the expansion valid above the resonance.
We have taken $\Delta m^{2}=2.5~10^{-3}~$eV$^{2}$.}
\end{figure}

\newpage

\begin{figure}[ht]
\begin{tabular}{cc}
\includegraphics[width=8cm,height=7cm]{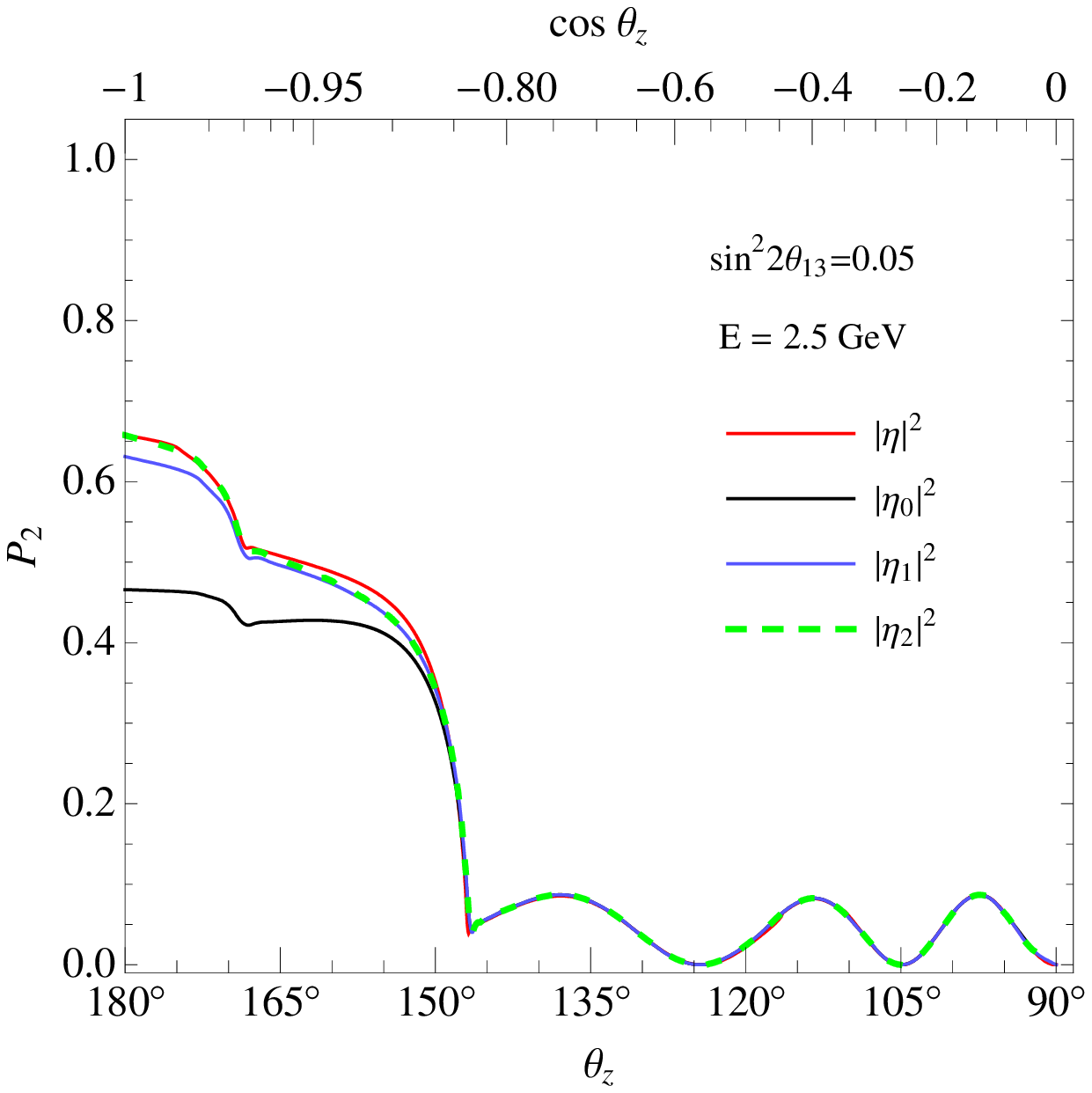}&
\includegraphics[width=8cm,height=7cm]{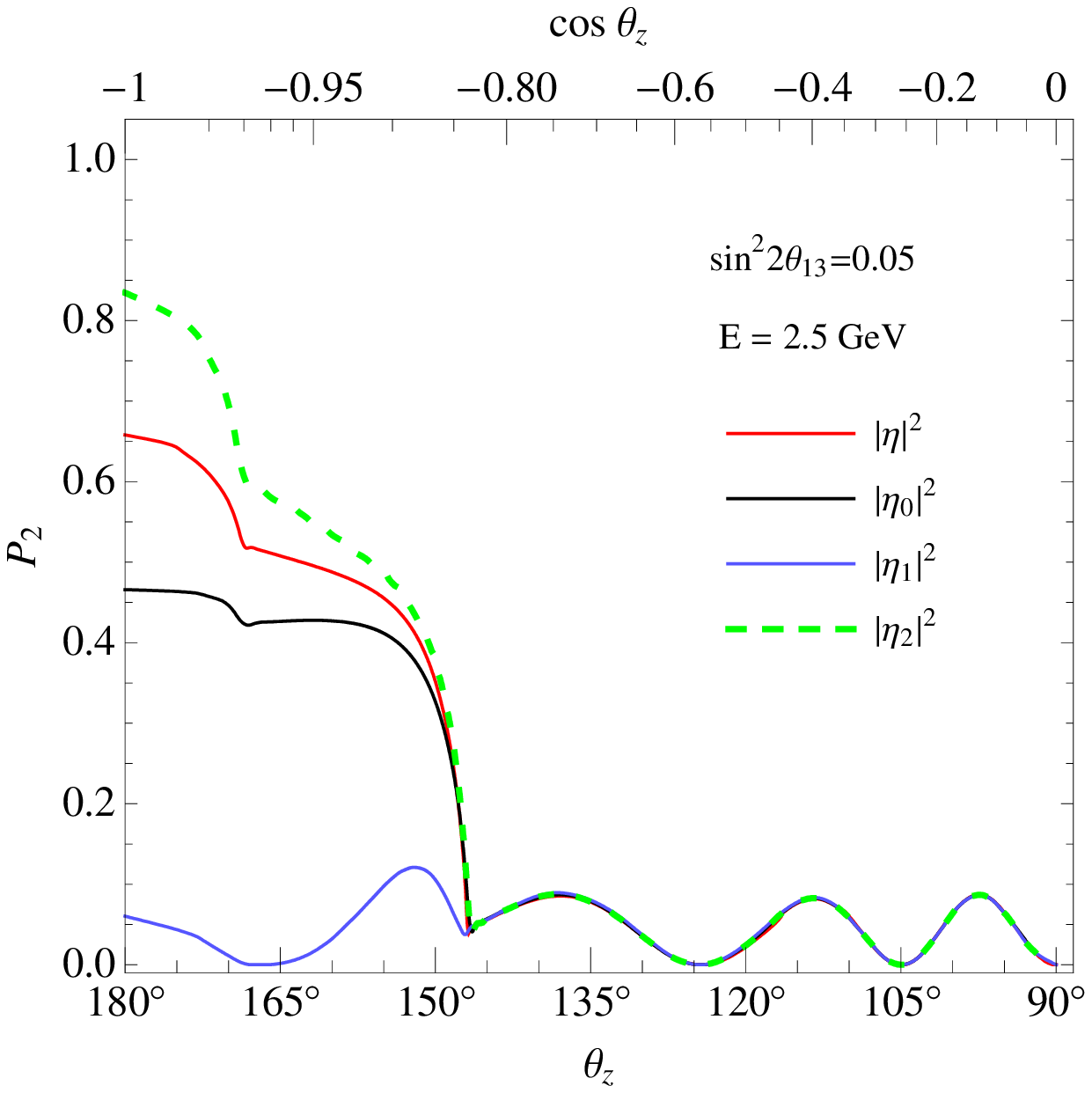}\\[1.5cm]
\includegraphics[width=8cm,height=7cm]{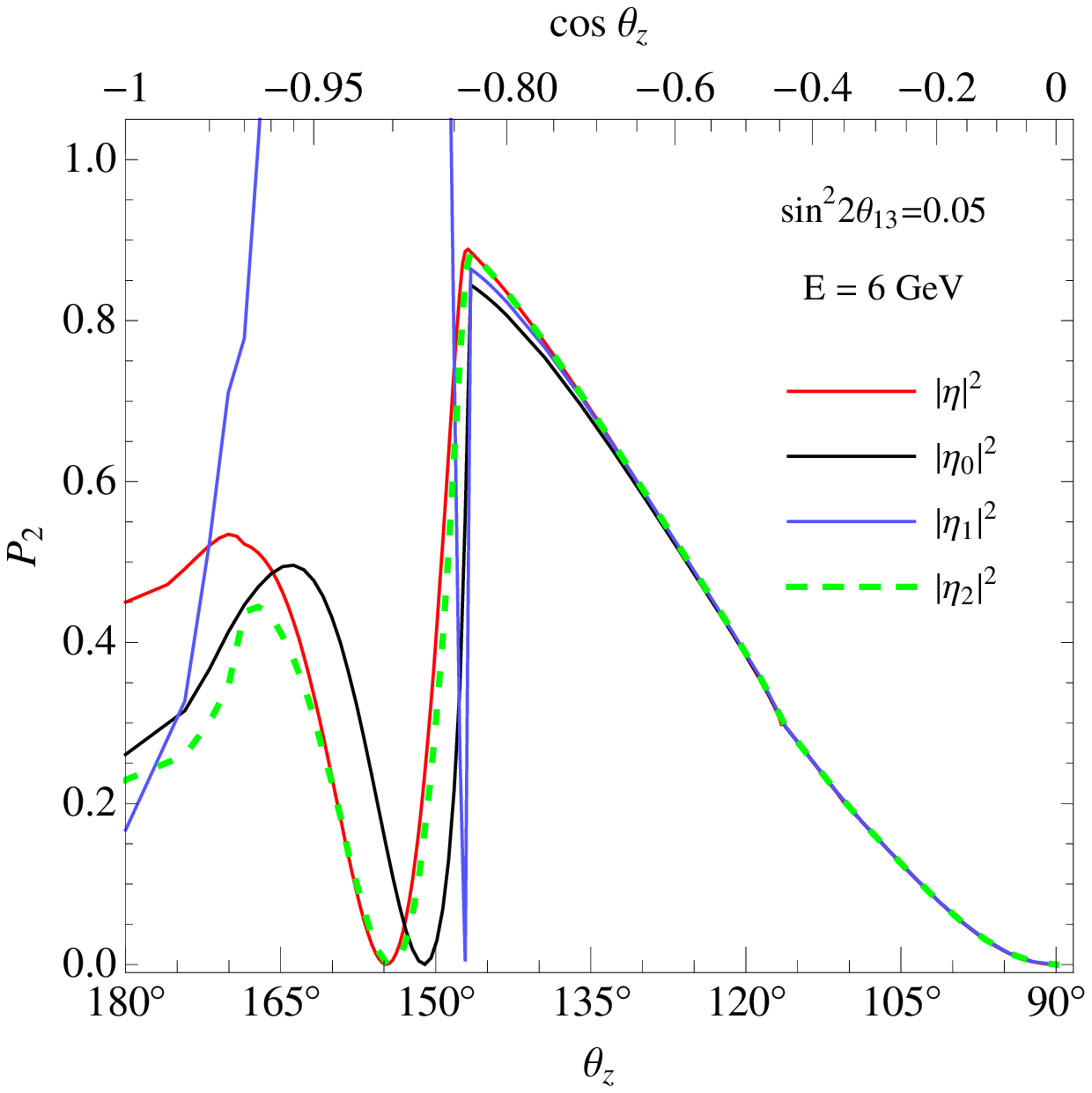}&
\includegraphics[width=8cm,height=7cm]{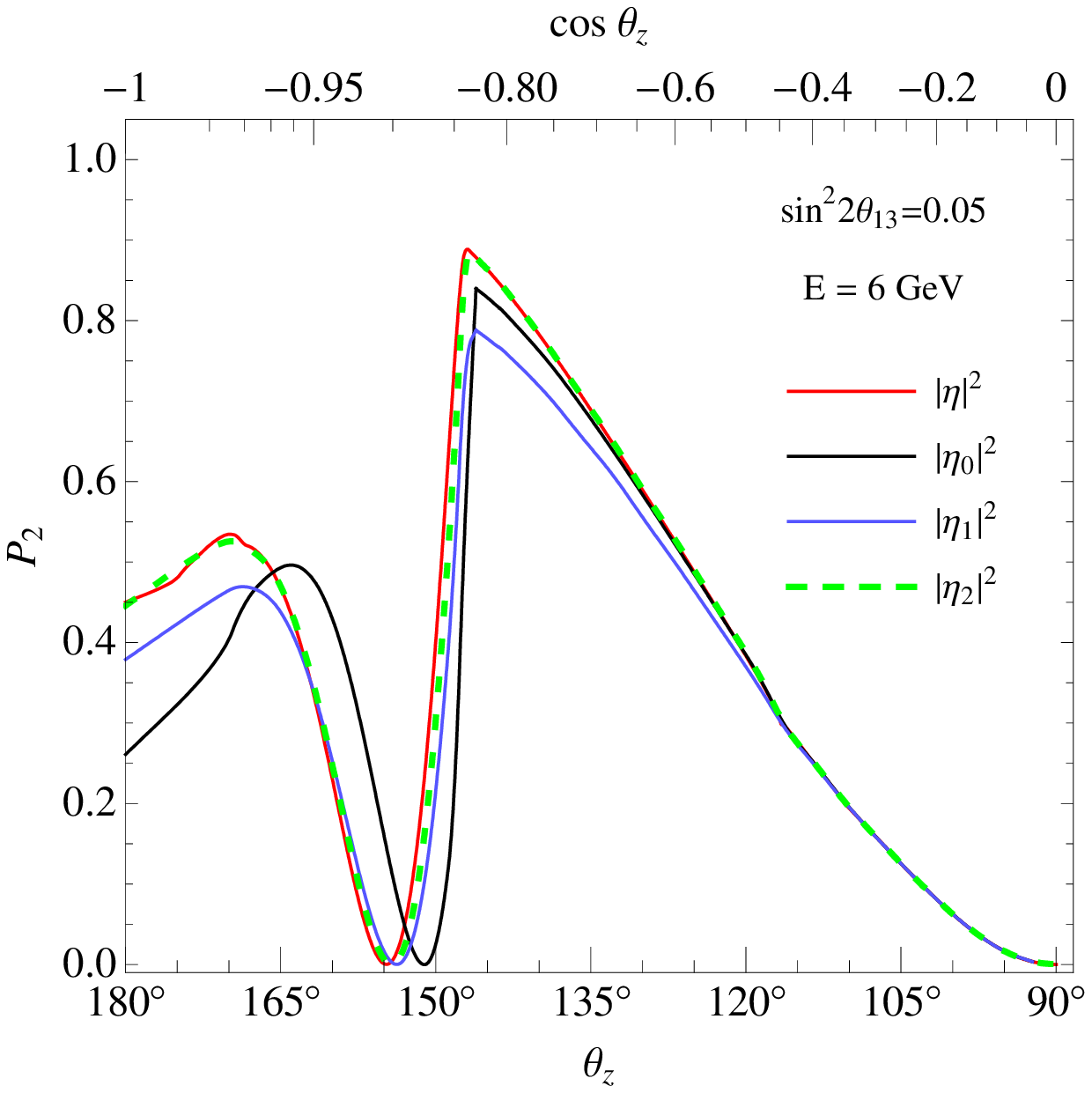}
\end{tabular}
\caption{\label{fig:Zenith3GeV} Oscillation probability $P_{2}$ versus 
the zenith angle $\theta_{z}$ for neutrinos propagating inside the Earth,  
for neutrino energies $E=2.5$ and 6 GeV. Left panels: probabilities 
obtained from the expansion valid below the MSW resonance, right 
panels: the same for the expansion valid above the resonance.
We have taken $\Delta m^{2}=2.5~10^{-3}~$eV$^{2}$.}
\end{figure}

\newpage

\begin{figure}[ht]
\begin{tabular}{cc}
\includegraphics[width=8cm,height=7cm]{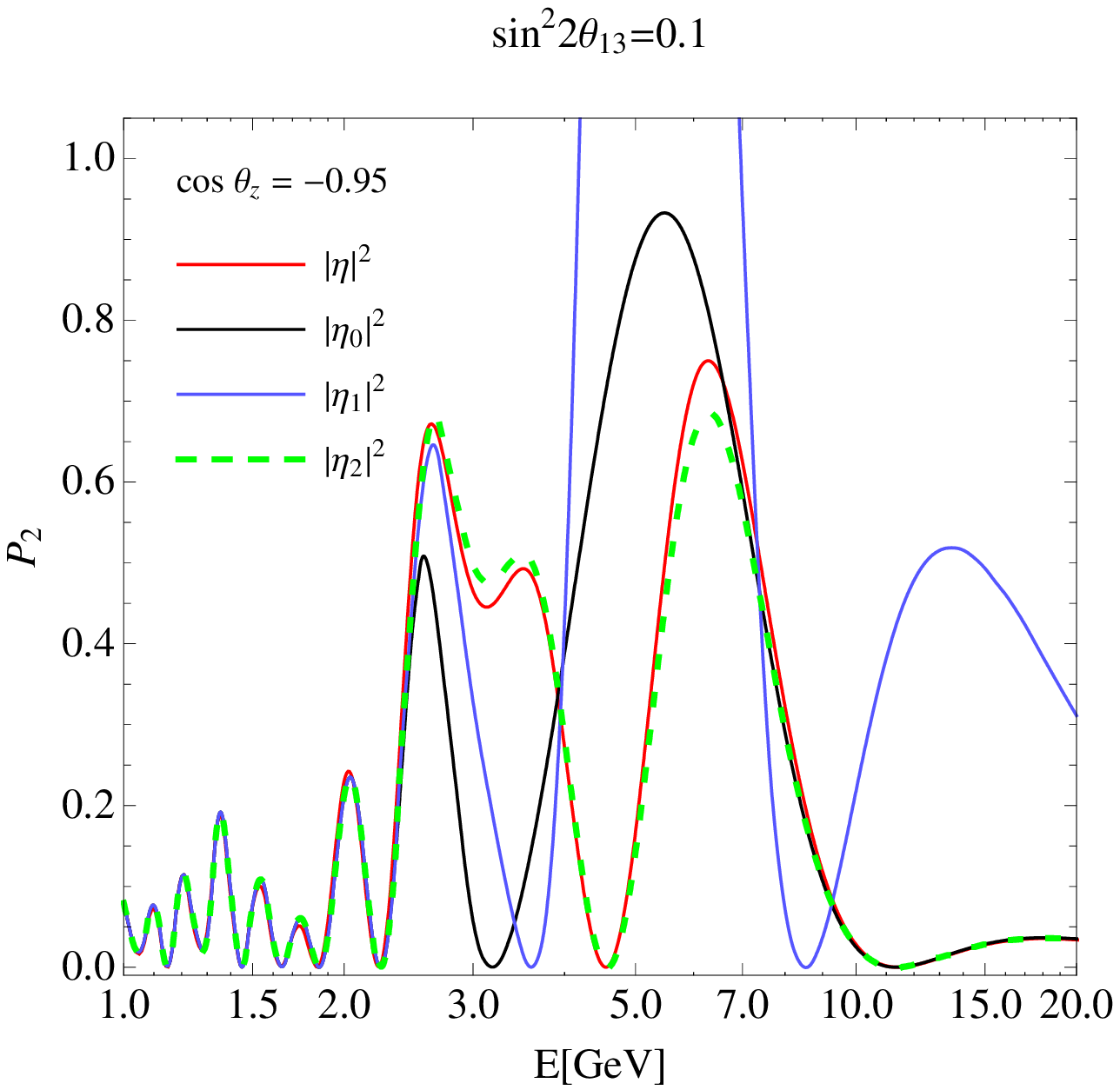}&
\includegraphics[width=8cm,height=7cm]{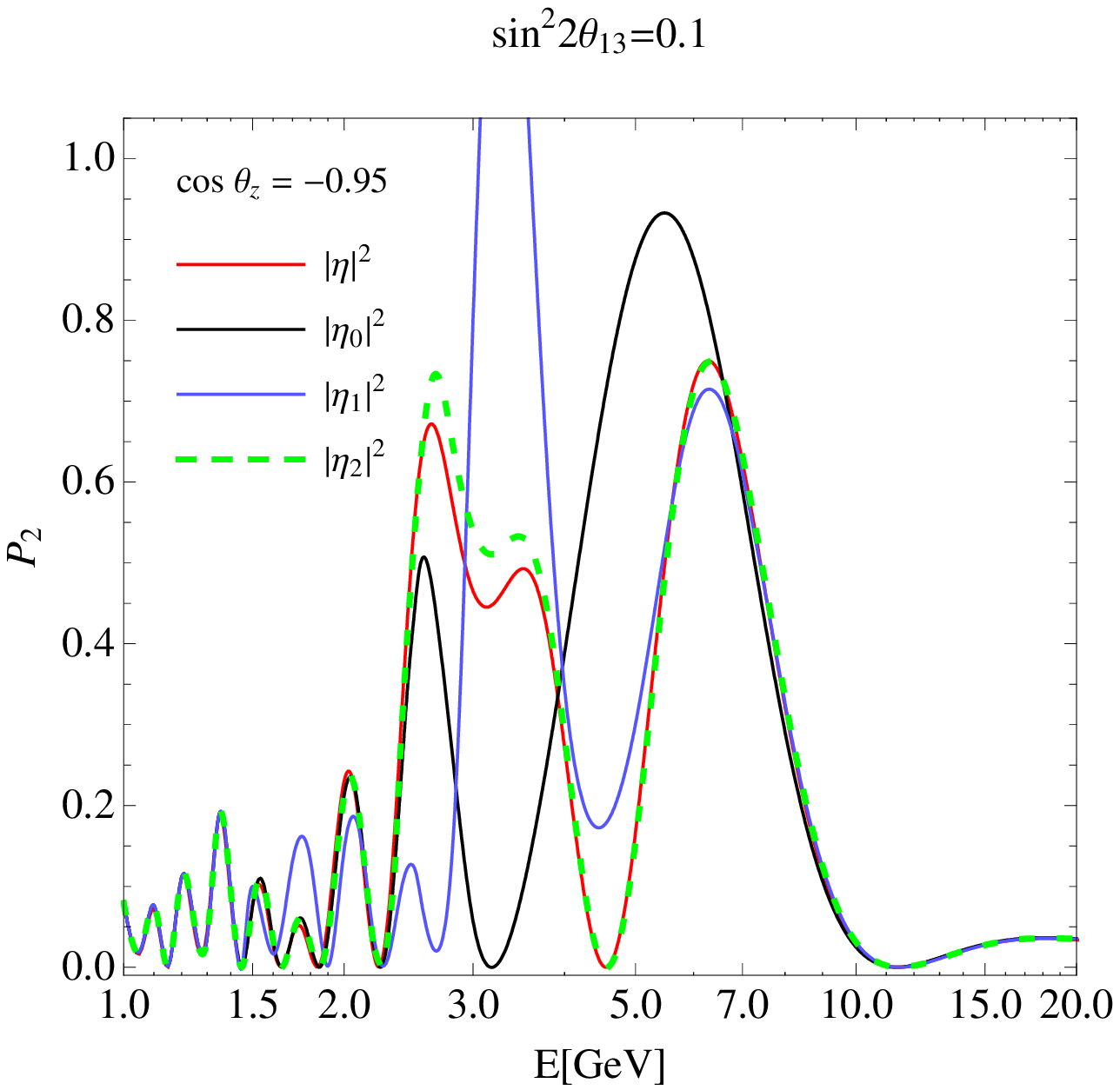}\\[1.5cm]
\includegraphics[width=8cm,height=7cm]{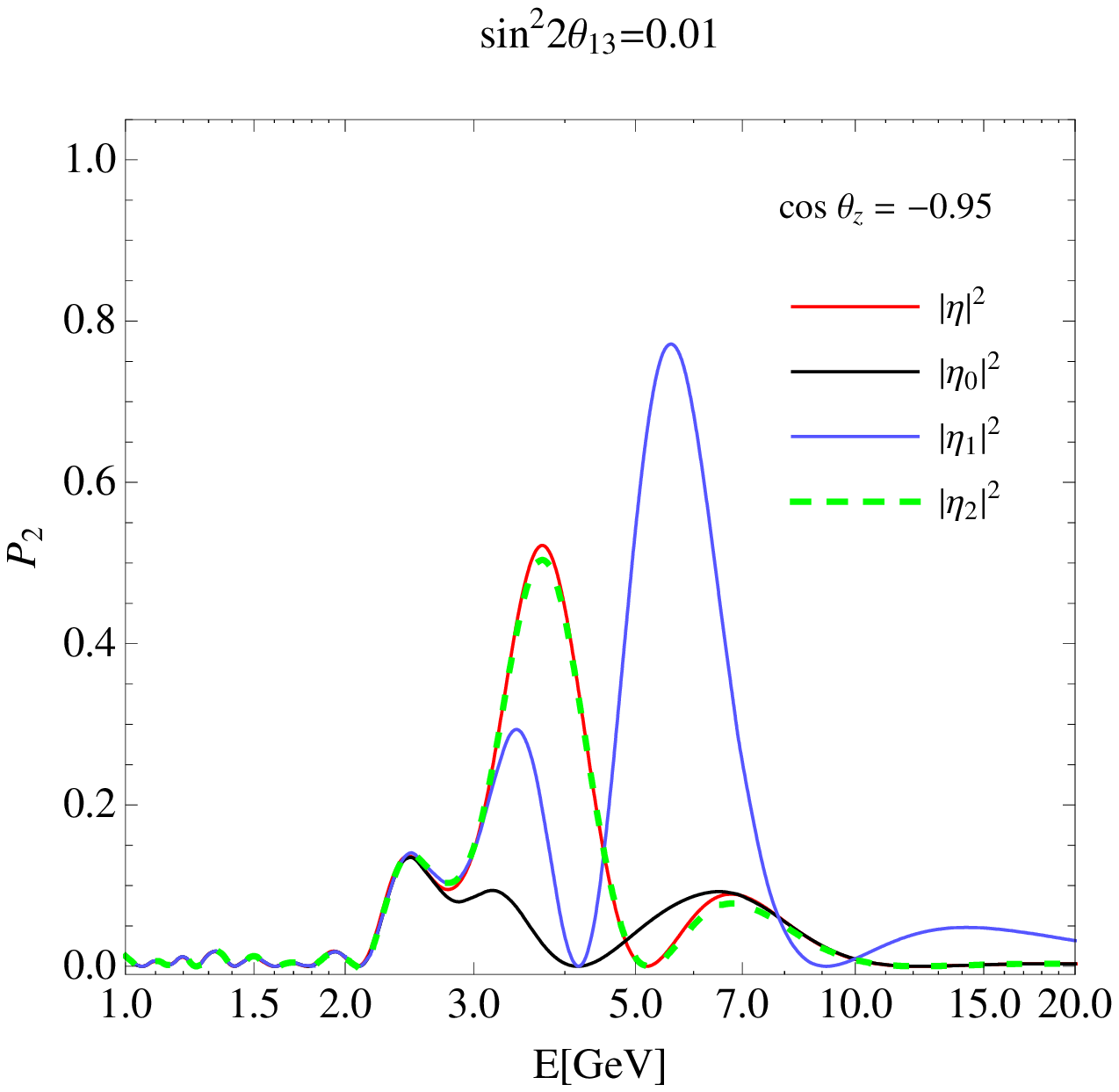}&
\includegraphics[width=8cm,height=7cm]{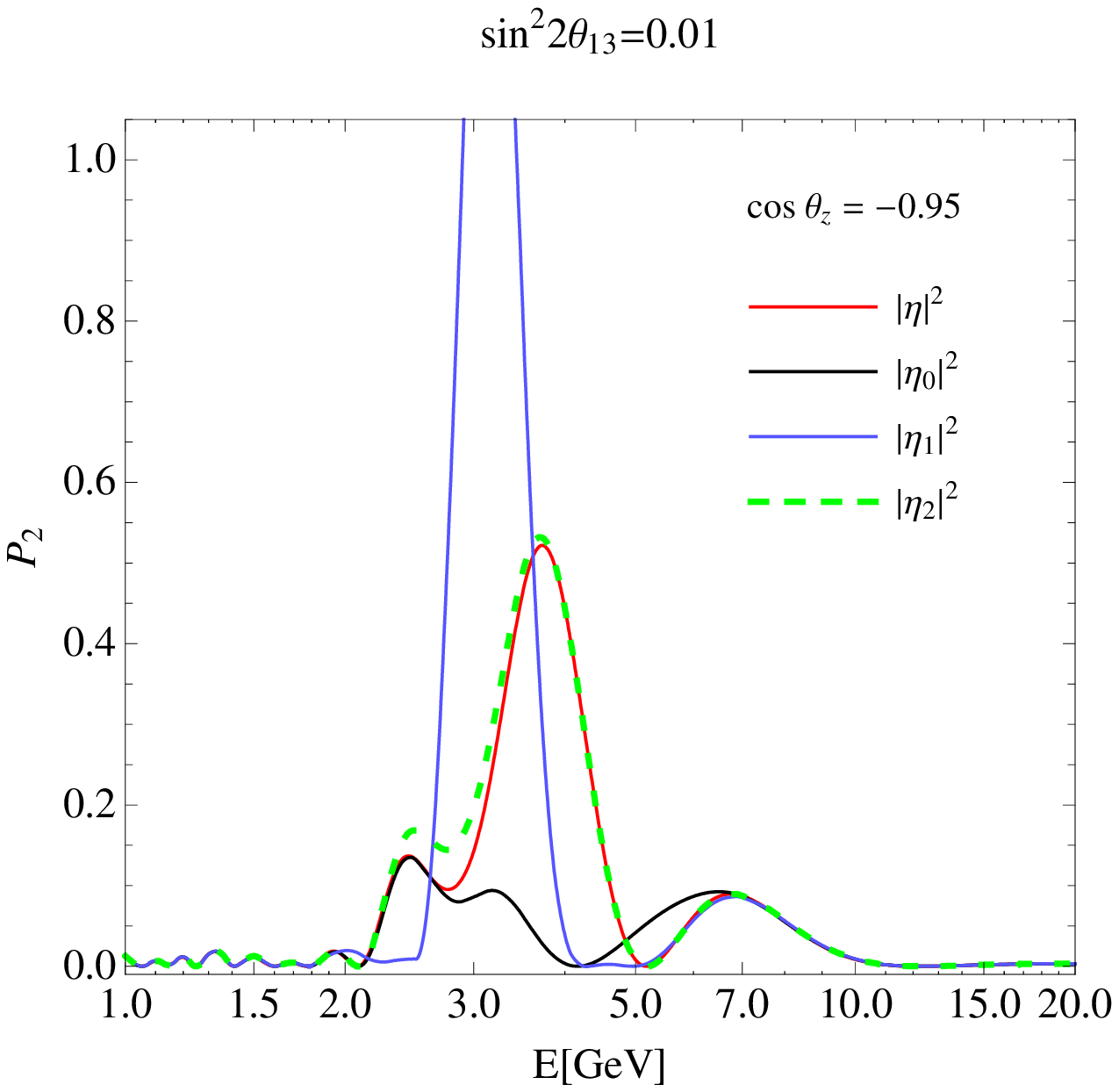}
\end{tabular}
\caption{\label{fig:sinvaluesbelow}Oscillation probability $P_{2}$ versus 
neutrino energy $E$ in the case of $\cos \theta_{z}=-0.95$ and for two 
different values of $\sin^{2} 2 \theta_{13}$. Left panels: probabilities 
obtained from the expansion valid below the MSW resonance, right 
panels: the same for the expansion valid above the resonance.  
We have taken $\Delta m^{2}=2.5~10^{-3}~$eV$^{2}$.}
\end{figure}

\newpage

\end{document}